\documentclass[aps,prd,notitlepage,reprint,twocolumn,showpacs,superscriptaddress,longbibliography,nofootinbib,floatfix]{revtex4-1}

\usepackage{amsmath,amssymb,amsfonts}
\usepackage{graphicx}
\usepackage{verbatim}
\usepackage{pifont}

\usepackage{hyperref}
\usepackage{slashed}
\usepackage{verbatim}
\newcommand{\be}{\begin{equation}}
\newcommand{\ee}{\end{equation}}
\newcommand{\bi}{\begin{itemize}}
\newcommand{\ei}{\end{itemize}}
\newcommand{\bea}{\begin{eqnarray}}
\newcommand{\eea}{\end{eqnarray}}

\newcommand{\ud}{\mathrm{d}}

\newcommand{\LCp}{{\scriptscriptstyle +}}

\newcommand{\LCperp}{{\scriptscriptstyle \perp}}
\newcommand{\para}{{\scriptscriptstyle \parallel}}

\usepackage[T1]{fontenc} 

\newcommand{\abar}{p}

\begin{document}
\title{Nonperturbative pair production in interpolating fields}

\author{Anton Ilderton}
\email[]{anton.ilderton@chalmers.se}
\affiliation{Dept.~Applied Physics, Chalmers University of Technology, SE-41296 Gothenburg, Sweden}

\author{Greger Torgrimsson}
\email[]{greger.torgrimsson@chalmers.se}
\affiliation{Dept.~Applied Physics, Chalmers University of Technology, SE-41296 Gothenburg, Sweden}

\author{Jonatan W\aa rdh}
\email[]{jontan@student.chalmers.se}
\affiliation{Dept.~Applied Physics, Chalmers University of Technology, SE-41296 Gothenburg, Sweden}

\begin{abstract}
We compare the effects of timelike, lightlike and spacelike one-dimensional inhomogeneities on the probability of nonperturbative pair production in strong fields. Using interpolating coordinates we give a unifying picture in which the effect of the inhomogeneity is encoded in branch cuts and poles circulated by complex worldline instantons. For spacelike inhomogeneities the length of the cut is related to the existence of critical points, while for lightlike inhomogeneities the cut contracts to a pole and the instantons become contractable to points, leading to simplifications particular to the lightlike case. We calculate the effective action in fields with up to three nonzero components, and investigate its behaviour under changes in field dependence.
\end{abstract}
\maketitle

\section{Introduction}
Pair production in strong fields gives us insight into nonperturbative fundamental physics~\cite{Sauter:1931zz,Heisenberg:1935qt,Schwinger:1951nm}. Advances in technology have in recent years driven interest in observing nonperturbative pair production using intense lasers~\cite{Schutzhold:2008pz,Dunne:2008kc,Dunne:2009gi,DiPiazza:2009py,Bulanov:2010ei,DiPiazza:2011tq,Gonoskov:2013ada,Hebenstreit:2014lra,Otto:2015gla}, which presents a veritable experimental challenge even with optimally focussed laser light~\cite{Gonoskov:2013ada}.

It has been found that the pair production probability tends to be higher (relative to the locally constant approximation) in time-dependent electric fields $E(t)$, and lower in position-dependent inhomogeneous electric fields $E(z)$~\cite{Dunne:2005sx}. In the latter case there is even a critical point beyond which the probability is identically zero~\cite{Nikishov:1970br,Gies:2005bz}. Between these two cases is an electric field with lightlike inhomogeneities, $E(t+z)$. In this case the probability is given exactly by the locally constant approximation~\cite{Tomaras:2000ag,Tomaras:2001vs}. 

Here we would like to understand more about how the spacetime dependence of an electric field affects the pair production probability. Note that the three cases above cannot be related by a Lorentz transformation. In order to investigate the transition between spacelike and timelike field inhomogeneities we will therefore consider electric fields depending on a single interpolating coordinate of the form $(1-\alpha) t + \alpha z$ for $\alpha\in[0,1]$, following~\cite{Hornbostel:1991qj,Ji:2001xd,Ji:2012ux}.

We will use the worldline formalism (see~\cite{Strassler:1992zr,Schubert:1996jj}, and e.g.~\cite{Dietrich:2013kza,Mansfield:2014vea,Edwards:2014bfa} for recent applications), and in particular worldline instantons~\cite{Affleck:1981bma,Schubert:1996jj,Dunne:2005sx,Dunne:2006st}. These are periodic, in general complex~\cite{Lavrelashvili:1989he,KeskiVakkuri:1996gn,Rubakov:1992az,Dumlu:2011cc}, solutions to the classical equations of motion. The classical action evaluated on these solutions, together with the contributions of fluctuations around the instantons, gives a semiclassical approximation of the effective action, and of the pair production probability. We will see that this contribution is an integral over the instanton itself. Since the instantons are complex, these contributions depend, by Cauchy's integral theorem, only on the structures which the instantons circulate in the complex plane.

This allows us to unify, and extend, previous observations on the impact of field inhomogeneities on pair production. We will see that for timelike and spacelike inhomogeneities the instantons circulate differently orientated branch cuts and so are fundamentally extended objects. As the field dependence becomes lightlike, though, the branch points coalesce and become poles, so that the instantons are contractable to points~\cite{Ilderton:2015lsa}. We will see that this leads to the known localisation of the effective action in the lightlike limit~\cite{Tomaras:2000ag,Tomaras:2001vs}.

Another motivation for this paper is recent interest in pair creation in two-component, rotating fields, which model the electric antinodes of a circularly polarised standing wave~\cite{CPL,Blinne:2013via,Strobel:2013vza,Strobel:2014tha}. We will here extend these calculations to electric fields depending on our interpolating coordinates, and with up to three components.

This paper is organised as follows. In the remainder of this introductory section we recall the worldline approach to pair production in strong fields, introduce our interpolating fields and outline the calculation to be performed. In Sect.~\ref{SECT:INST} we present an explicit example of the instantons in an interpolating Sauter pulse, and relate the field inhomogeneity to the structures circulated by the instanton.  In Sect.~\ref{SECT:GAMMA} we complete the calculation of the effective action in electric fields with up to three nonzero components depending on the interpolating coordinate. In Sect.~\ref{SECT:PRAT} we analyse the behaviour of the effective action as a function of the interpolating parameter in several examples. We conclude in Sect.~\ref{SECT:CONC}.

\subsection{Pair creation in interpolating fields}
The probability $\mathbb P$ of pair production in an external field is related to the effective action $\Gamma$ by
\be
	\mathbb{P}_\text{pairs}=1-e^{-2\text{Im }\Gamma} \;.
\ee
Our starting point is the one-loop worldline representation of $\Gamma$ in terms of an integral over proper time $T$ and a path integral over periodic paths $x^\mu$,
\be\label{start Gamma}
	\Gamma=\int\limits_0^\infty\frac{\ud T}{T}\oint\mathcal{D}x\ e^{-i S} \;,
\ee
where the classical action is
\be\label{S1}
	S=\frac{m^2T}{2}+\int\limits_0^1\!\ud\tau\ \bigg[\frac{\dot{x}^2}{2T}+ e A_\mu(x)\dot{x}^\mu \bigg]\;.
\ee
Here $\tau$ parameterises the worldline. Our first aim is to examine how $\Gamma$ depends on fields of given strength, shape and direction as the field dependence, or inhomogeneity, varies from temporal to spatial. To do so we introduce interpolating coordinates $\{q,d\}$ as~\cite{Hornbostel:1991qj,Ji:2001xd,Ji:2012ux}
\be\label{qd-def}
	\begin{pmatrix} q \\ d \end{pmatrix} 
	=
	\begin{pmatrix} \cos\frac{\theta}{2} & \sin\frac{\theta}{2} \\ -\sin\frac{\theta}{2} & \cos\frac{\theta}{2} \end{pmatrix} 
	\begin{pmatrix} t \\ z \end{pmatrix} \;,
\ee
in which $\theta \in [0,\pi]$. This is not a Lorentz transformation and so we are considering Lorentz inequivalent cases.  We consider electric fields of a given profile $E^3(q)$ (i.e.~given form and amplitude) which always point in the $z$-direction. As $\theta$ varies we interpolate between time-dependent homogeneous electric fields $E^3(t)$ at $\theta=0$, fields depending on lightfront time $(t+z)/\sqrt{2}$ at $\theta=\pi/2$ and finally spatially inhomogeneous static electric fields $E^3(z)$ at $\theta=\pi$. (The coordinate $d$ interpolates between position $z$, lightfront position $(-t+z)/\sqrt{2}$, and time $-t$.) 

Let $A_\para'(q) = E^3(q)$, then we take as gauge potential
\be\label{gauge val}
	A_\mu(q) = A_\para(q) \hat{d}_\mu \;,
\ee
where $\hat{d}.x = d$. It is easily checked that only $F_{03} = E^3(q)$ is nonzero. 

We will also consider a time-dependent rotating electric field under the transformation (\ref{qd-def}).  For this we use the gauge potential, $\perp=\{1,2\}$,
\be\label{gauge val2}
	A_\mu(q) = \delta_\mu^\LCperp A_\LCperp(q) \;,
\ee
The transverse potential gives electric and magnetic fields
\be\label{EB-trans}
	E^i(q) = \cos\tfrac{\theta}{2} A'_i(q) \;, \quad B^i(q) = -\sin\tfrac{\theta}{2} \epsilon^{ij} A'_j(q) \;.
\ee
For $\theta=0$ (\ref{EB-trans}) describes a time-dependent, rotating electric field, for $\theta=\pi/2$ a plane wave, and for $\theta=\pi$ a static, inhomogeneous magnetic field.

Note that the amplitude of the {\it transverse} fields transforms as we rotate. Nevertheless we present the calculation of the effective action in the two types of field together, i.e.~we will consider three-component electric fields, with the `longitudinal' (strictly, $z$) and transverse components described by (\ref{gauge val}) and (\ref{gauge val2}) respectively, the potential being simply the sum of these,
\be
	A_\mu\dot{x}^\mu=A_\para(q)\dot{d}+A_\LCperp(q)\dot{x}^\LCperp \;.
\ee
To calculate $\Gamma$ we follow essentially the same steps as in~\cite{Dunne:2006st}. We first find the periodic instanton solutions of the classical equations of motion in order to identify the dominant, exponential, contributions to $\Gamma$. Once we have the instantons, we evaluate their classical action, which is the saddle point contribution to $\Gamma$. We then integrate over fluctuations around the instantons in order to obtain a prefactor contribution. Finally we perform the $T$--integral, also with a saddle point approximation, in order to obtain the final expression for~$\Gamma$. This calculation will reveal new insights into the structure and properties of complex instantons. Regarding this, for the symmetric fields ($E(t)= E(-t)$) usually considered, the exponent in (\ref{start Gamma}) becomes real after rotating both proper time $T$ and time $t$ to Euclidean space. However, this rotation does not make the exponent real for general fields, so the instantons will in general be complex~\cite{Lavrelashvili:1989he,KeskiVakkuri:1996gn,Rubakov:1992az,Dumlu:2011cc}. We will therefore use Minkowski coordinates throughout.

Our results will be valid in the semiclassical regime, i.e.~for weak fields and not too large adiabaticity~$\gamma$,
\be\label{adi0}
	\gamma = \frac{m\omega}{eE_0} \;,
\ee
where $E_0$ and $\omega$ are typical strength and frequency scales of the considered field. This can be understood by rescaling $T\to T/(eE_0)$ and $x\to x/\omega$ in (\ref{S1}), which makes the whole action inversely proportional to $E_0$ and the integral term inversely proportional to $\gamma^2$. See~\cite{Dunne:2005sx,Kim:2007pm} for more details.

We now set $m=1$ and absorb factors of $e$ into the gauge potentials, reinstating the mass only in final results. Throughout we use the notation
\be
	c:= \cos\theta \;, \qquad s:=\sin\theta \;.
\ee
%

\section{Complex worldline instantons in interpolating fields\label{SECT:INST}}
In our interpolating coordinates the equations of motion are
\be\label{Lorentz}
\begin{split}
	c\ddot{q}-s\ddot{d}&= TA'_\mu\dot{x}^\mu \;, \\
	c\ddot{d}+s\ddot{q}&=TA'_\para\dot{q} \;, \\
	\ddot{x}^\LCperp &=TA'_\LCperp\dot{q} \;,
\end{split}
\ee
which are just the Lorentz force equations. The latter two equations are readily solved for $d$ and $x^\LCperp$ in terms of~$q$,
\bea
\label{dd}
\dot{d} &=& \frac{1}{c}\Big[T(A(q)-p)_\para-s\dot{q}\Big] \;, \\
\label{dxperp}
\dot{x}^\LCperp &=& T(A(q)-p)_\LCperp \;,
\eea
where the integrations constants $p_{\para,\LCperp}$ are determined by integrating (\ref{dd}) and (\ref{dxperp}); periodicity of the instantons requires that $p_{\para,\LCperp}$ be equal to the $\tau$-average of the potential,
\be
	p_{\para,\LCperp}=\int\limits_0^1\ud\tau A_{\para,\LCperp}(q) =: \langle A\rangle_{\para,\LCperp}\;.
\ee
This means that $\abar$ depends, in general, on $c$, $\gamma$ and proper time~$T$~\cite{CPL}.  We also find
\be\label{xd2}
	\dot{x}^\mu \dot{x}_\mu = \text{const} =: T^2 a^2 \;,
\ee
which defines a fourth constant $a$, see below. Using \eqref{dd}, \eqref{dxperp} and \eqref{xd2} yields an expression for $\dot{q}$ as a function of~$q$;
\be\label{dq}
	\dot{q}^2=T^2\big(ca^2+(A-p)_\para^2+c(A-p)_\LCperp^2\big) \;.
\ee
To understand what $p$ represents it is helpful to rewrite the equation of motion for $q(\tau)$ as
\be\label{ddp}\begin{split}
	\ddot{q}=T^2\big(A'_\para(A-p)_\para+cA'_\LCperp(A-p)_\LCperp\big) \;.
\end{split}
\ee
and then to compare with the {\it phase-space} equations of motion, see e.g.~\cite{Dumlu:2011cc}. In our interpolating coordinates the phase-space equation of motion for $q(\tau)$ is precisely (\ref{ddp}) with $\abar_{\para,\LCperp}$ appearing as the canonical momenta (conjugate to $\hat{d}.x$ and $x^\LCperp$ respectively) of the produced electron-positron pair.  We will show below that the values of $\abar_{\para,\LCperp}$ required by periodicity correspond to the saddle point of WKB integrals over $p_{\para,\LCperp}$.  Note that $A$ and $\abar$ appear only in the gauge invariant combination $A-\abar$, which obeys $\langle A - \abar \rangle = 0$~\cite{Ilderton:2015lsa}. 

The solutions to (\ref{dq}), and to the other equations of motion, are in general complex, i.e.~closed curves in the complex plane. Taking a square root of (\ref{dq}) implies that the velocity $\dot{q}$ has a branch cut in the complex $q$-plane, with the branch points corresponding to the `turning points' where $\dot{q}=0$.  It is at this stage useful to consider examples of the instanton solutions, before going on to the calculation of the effective action.

\subsection{The Sauter pulse}
For our example we turn off the transverse fields, and take for the longitudinal electric field the well-studied Sauter pulse
\be\label{Sauter}	
	E^3(q) = E_0\, \text{sech}^2(\omega q) \;.
\ee    
The shape and strength of this field do not change under the transformation (\ref{qd-def}), so here we examine purely the dependence on the spacetime inhomogeneity.

The instanton solutions to (\ref{Lorentz}) are
\bea
\label{psech}
	&q(\tau)& = \frac{1}{\omega}\sinh^{-1}\bigg[\frac{i \bar\gamma\sqrt{c}}{\sqrt{1+{\bar\gamma}^2c}}\sin 2n\pi(\tau - \tau_0) \bigg] \;, \\
\label{nsech}
	&d(\tau)& = - \frac{s}{c}\, q(\tau) + \\
\nonumber &&-\frac{1}{c\omega} \frac{1}{\sqrt{1+\bar\gamma^2c }}\;\sinh^{-1}\bigg[\bar\gamma\sqrt{c} \cos2n\pi(\tau-\tau_0)\bigg],
\eea
where $\gamma$ is the adiabacity parameter from (\ref{adi0}), $\bar\gamma = a \gamma$ with $a$ as in (\ref{xd2}), $\tau_0$ is a constant and $n\in\mathbb{Z}$ is the turning number of the instanton. The constant $a$ is related to proper time $T$ via the relation
\be\label{Ta}
	eE_0 T\sqrt{1+\bar\gamma^2 c} = -2n \pi i \;.
\ee
As a check, we note that (\ref{psech})--(\ref{Ta}) agree with the solutions in~\cite{Dunne:2006st} in the limit\footnote{Our notation differs from that in~\cite{Dunne:2006st}, as both $t$ and $T$ are there rotated to Euclidean space. Also our $T$ is a factor of 2 larger.} $c\to1$. 

Observe first that a real $\tau_0$ can be absorbed into a redefinition (reparameterisation) of $\tau$. In this case we obtain a purely {\it imaginary} $q(\tau)$ and a {\it complex} $d(\tau)$.  The instanton $q(\tau)$ oscillates back and forth along a straight line between the two turning /branch points where $\dot{q}$ vanishes; using (\ref{psech}) we have:
\be\label{SECH-VEL}
	\dot{q}^2= T^2\bigg( c a^2+\frac{1}{\gamma^2} \tanh^2(\omega q) \bigg) \;.
\ee

\begin{figure}[t!]
	\centering\includegraphics[width=0.6\columnwidth]{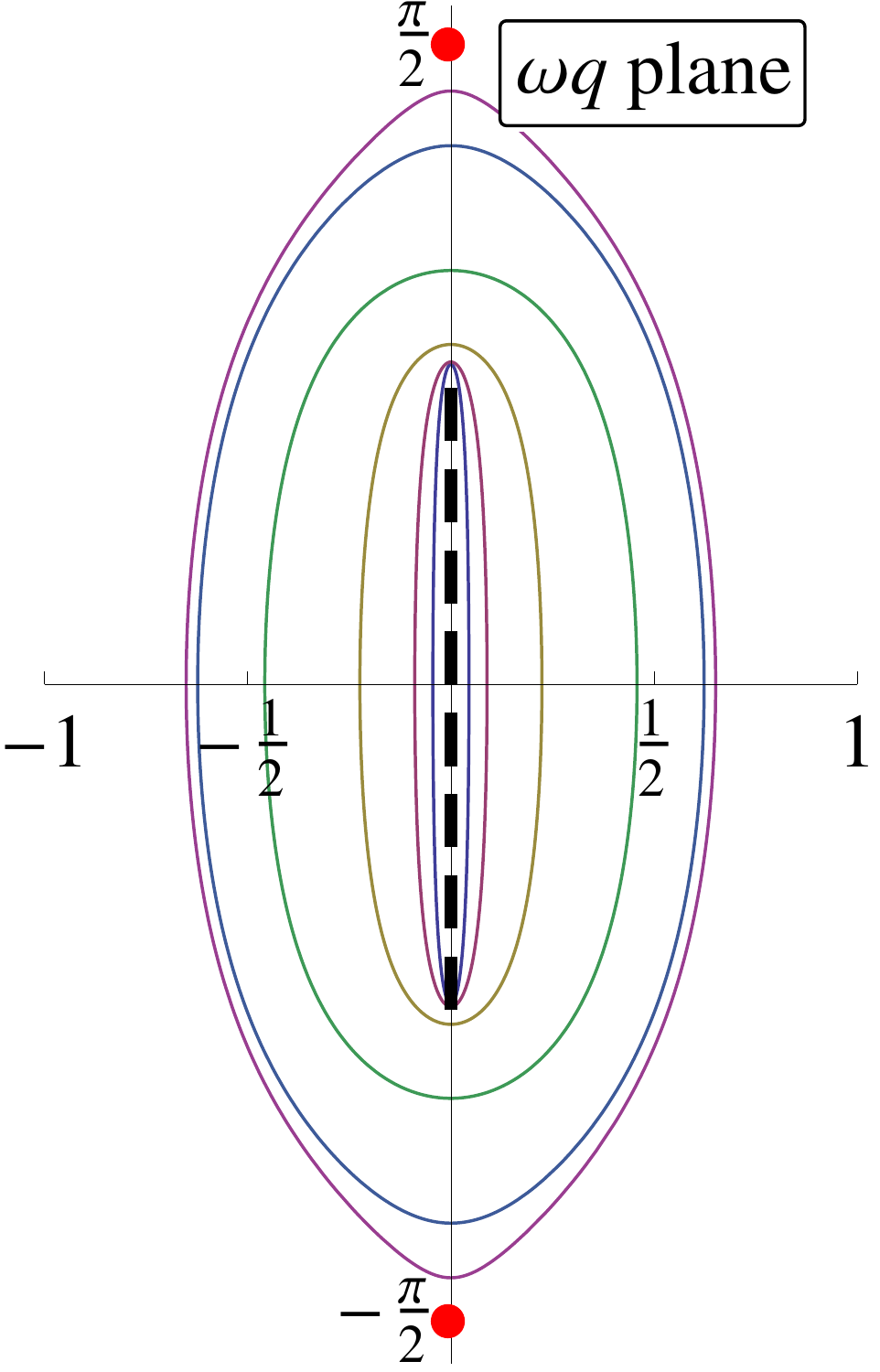}
	\caption{\label{FIG:P} The complex instantons $q(\tau)$ in a time-dependent $\text{sech}^2$ electric field, $a=c=\gamma=1$ and various imaginary $\tau_0$. As $\text{Im }\tau_0\to 0$ the instantons become purely imaginary (purely real in Euclidean space) and sit on the branch cut, shown as the dashed/black line. The instantons cannot cross the poles in the field at $\pm i\pi/2$, shown as dots.}
\end{figure}
Consider then an imaginary $\tau_0$. This gives a complex instanton $q(\tau)$ that forms a loop {\it around} the turning points in the complex $q$ plane~\cite{Froman,Bender:1977dr,Kim:2007pm,Ilderton:2015lsa}. As the imaginary part of $\tau_0$ becomes larger, the size of the instanton loop increases until it encounters poles in the potential at $\omega q = \pm i \pi/2$. There are no periodic solutions beyond this point; similar behaviour is seen in dynamically assisted pair production schemes~\cite{Schutzhold:2008pz,Schneider:2014mla,Linder:2015vta}.

\begin{figure}[t!]
	\includegraphics[width=0.9\columnwidth]{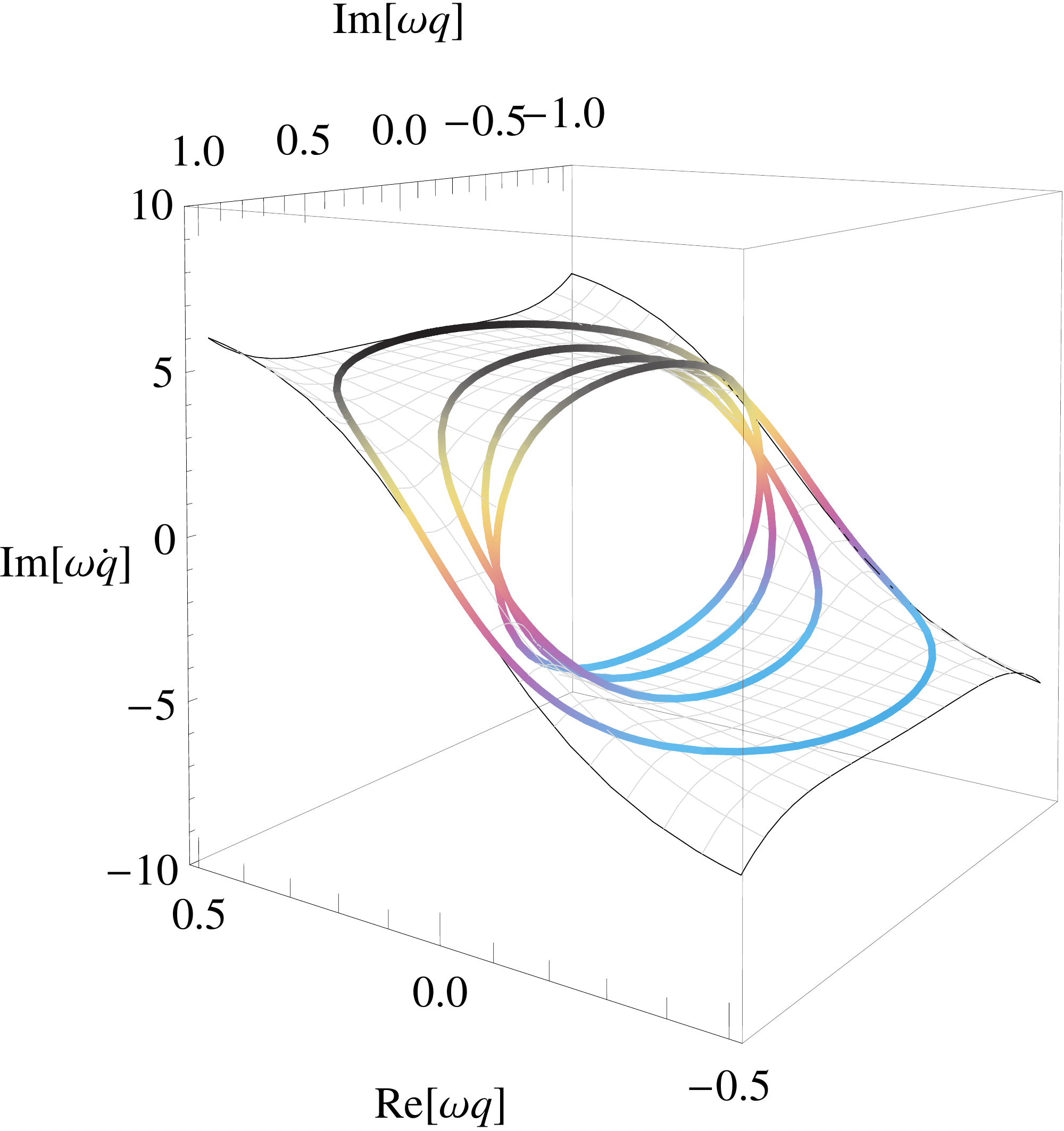}
5	\caption{\label{FIG:3D} The complex instantons $q(\tau)$ from Fig.~\ref{FIG:P}, plotted at a height of $\text{Im }\omega \dot{q}$ using (\ref{SECH-VEL}), in order to to show how they circle the branch cut in the instanton velocity $\dot{q}$.}
\end{figure}

In Fig.~\ref{FIG:P} we plot the instantons $q(\tau)$ for $c=\gamma=1$, various $\tau_0$, and $a=1$; the latter gives, as we will see below, the dominant contribution to $\Gamma$. Both the small and large (imaginary part of) $\tau_0$ behaviours described above can be seen. In Fig.~\ref{FIG:3D} we plot the same instantons but in three dimensions, using $\text{Im }\dot{q}$ as the third dimension; this shows how the instantons circulate the branch cut in the velocity, and contract around it as the imaginary part of $\tau_0$ goes to zero.

\begin{figure}[t!]
	\centering\includegraphics[width=\columnwidth]{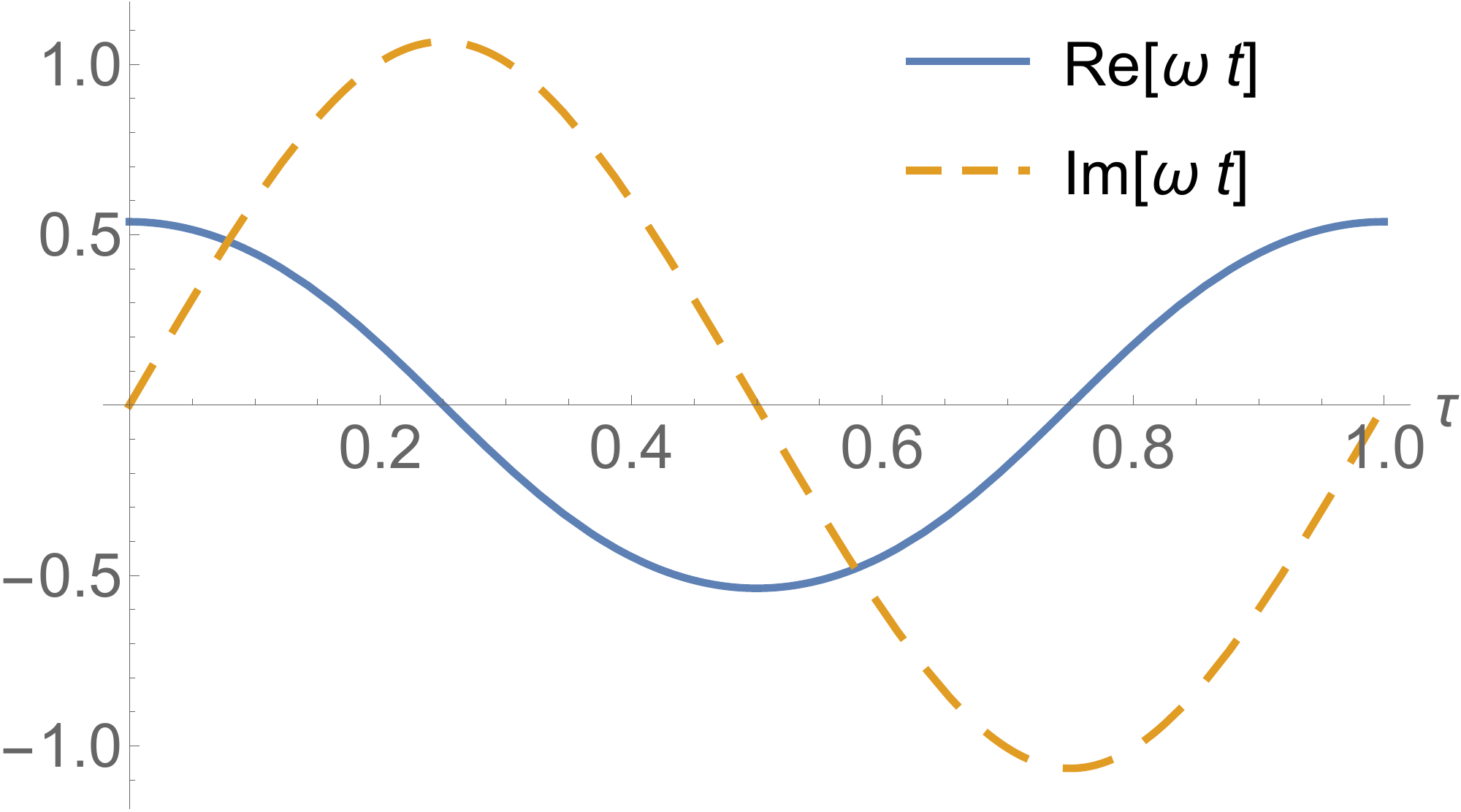}
	\centering\includegraphics[width=\columnwidth]{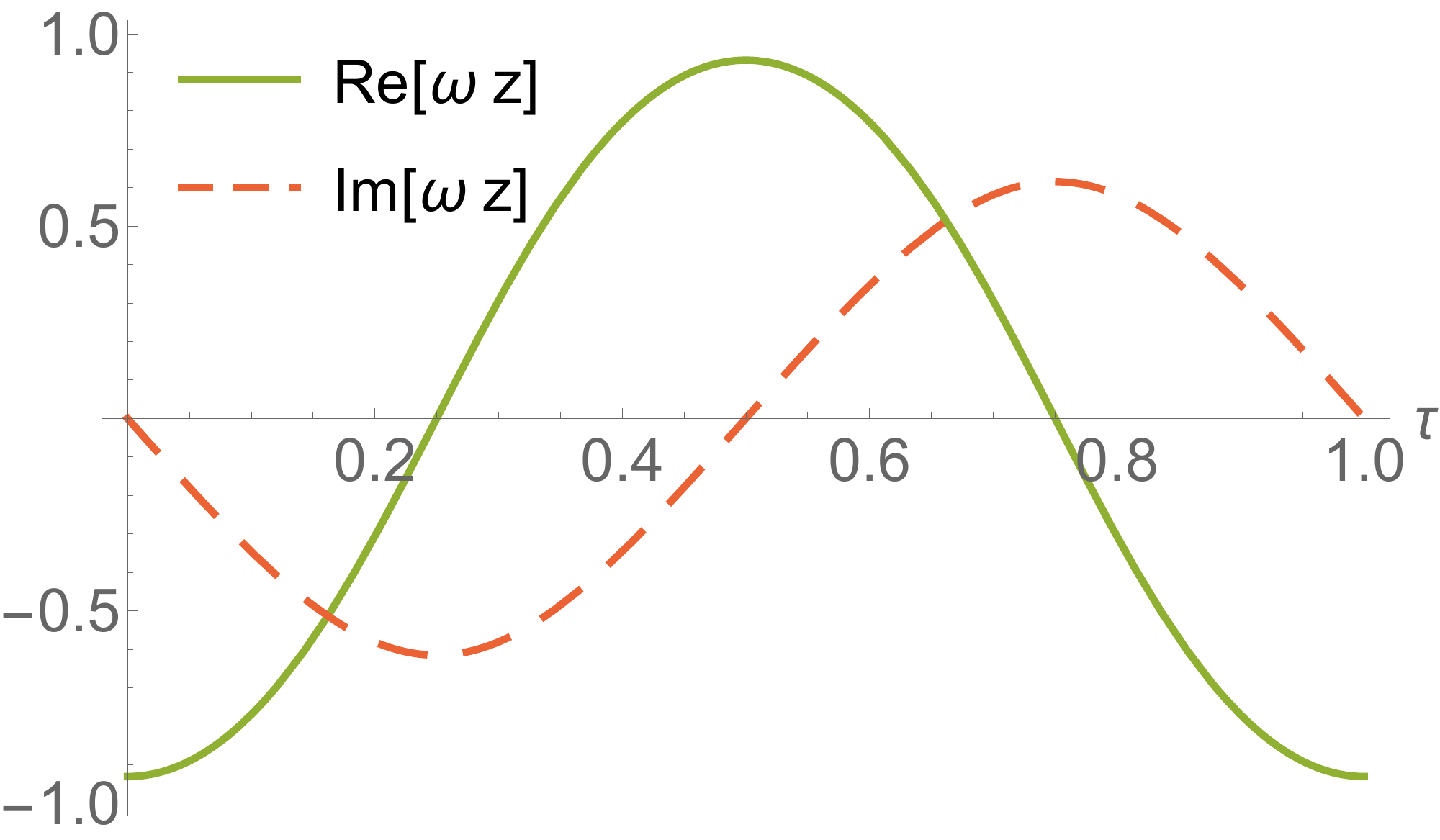}
	\caption{\label{FIG:N} The instantons $t(\tau)$ and $z(\tau)$ for $c=1/2$, i.e.~an electric field $E(t+z\sqrt{3})$, and $\gamma=a=1$. Even for $\tau_0=0$ the instantons have both real and imaginary parts.}
\end{figure}

These complex instantons generalise the Euclidean-real solutions in~\cite{Dunne:2006st}.  Very similar structures to those found here are seen in complexified classical motion~\cite{Bender:2009jg} and in the WKB/phase-integral formalism~\cite{Kim:2007pm,Froman}. We remark that in the case of time-dependent fields, i.e.~$c\to 1$, then $q\to t$ is purely imaginary for $\tau_0=0$, while $d\to z$ is real. This would imply an entirely real instanton {\it after} rotating $t$ to Euclidean space. For general $c$, though, we have that
\be
	t =  q \cos \tfrac{\theta}{2}-d \sin \tfrac{\theta}{2} \;, \quad z = q \sin \tfrac{\theta}{2} + d \cos \tfrac{\theta}{2} \;,
\ee
so that even for $\tau_0=0$ both $t$ and $z$ will in general have both real and imaginary parts, even after rotating to Euclidean space. See Fig.~\ref{FIG:N} for an example.

\begin{figure}[t!!!]
\includegraphics[width=.53\columnwidth]{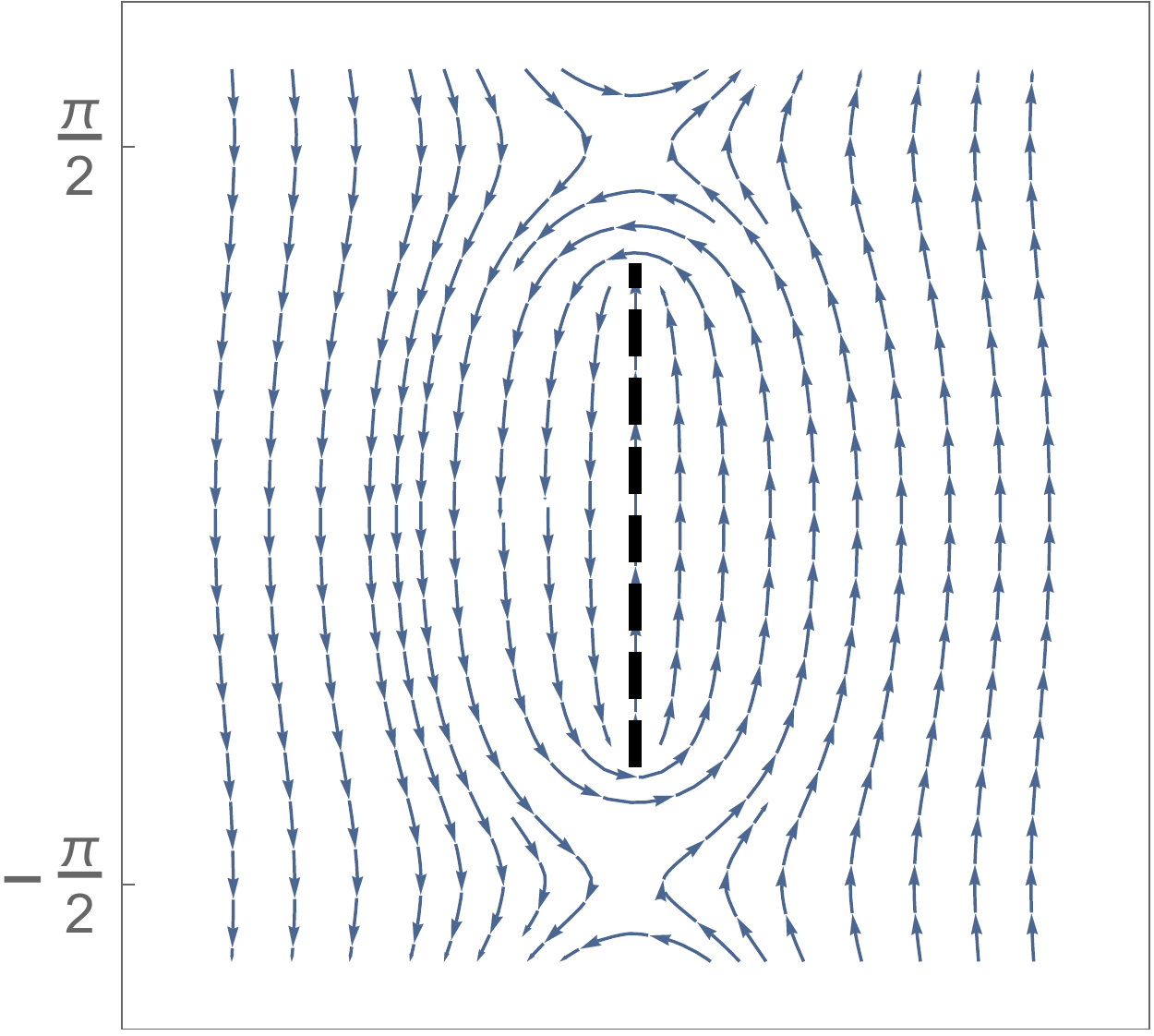}
\includegraphics[width=.53\columnwidth]{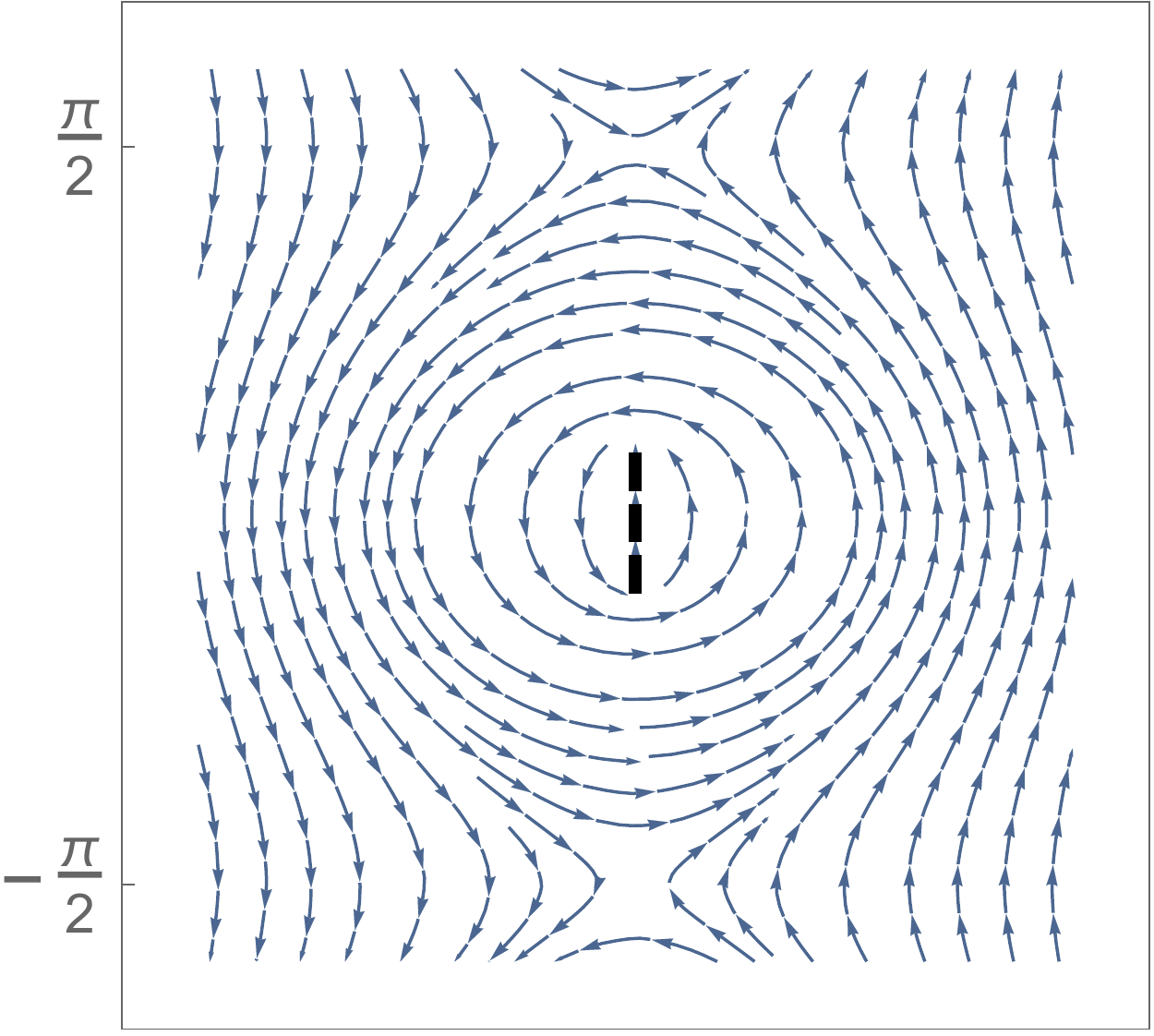}
\includegraphics[width=.53\columnwidth]{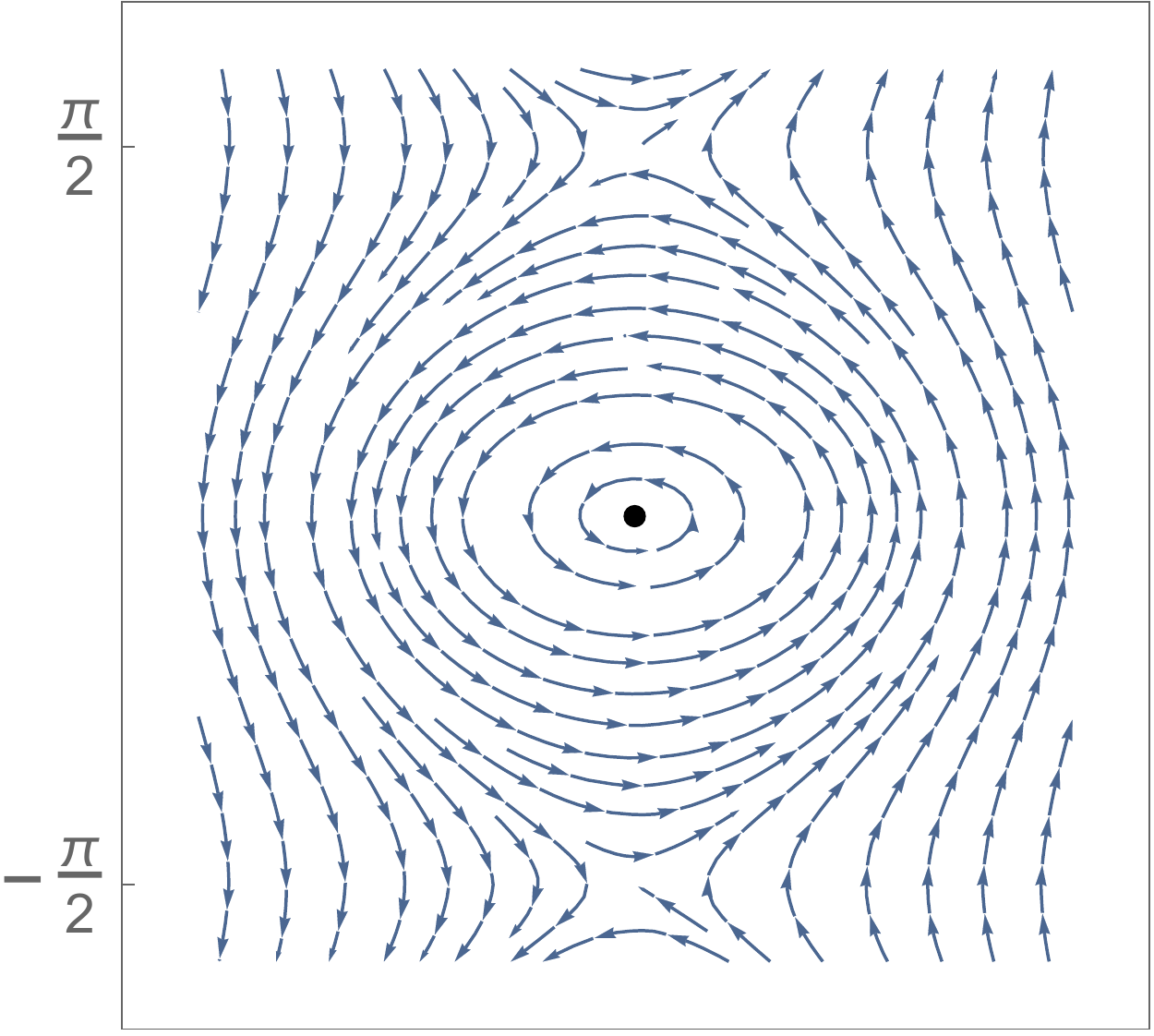}
\includegraphics[width=.53\columnwidth]{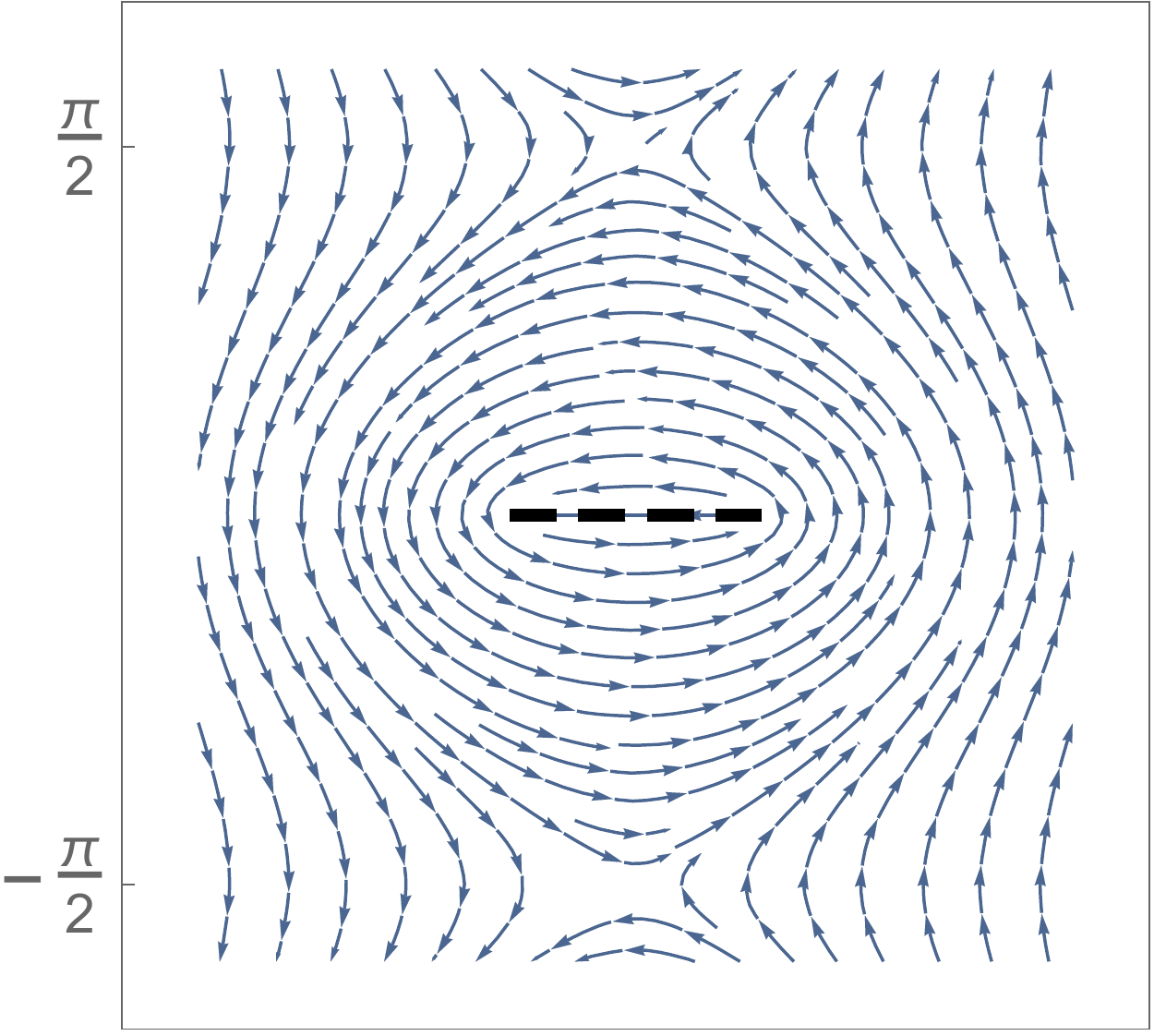}
\includegraphics[width=.53\columnwidth]{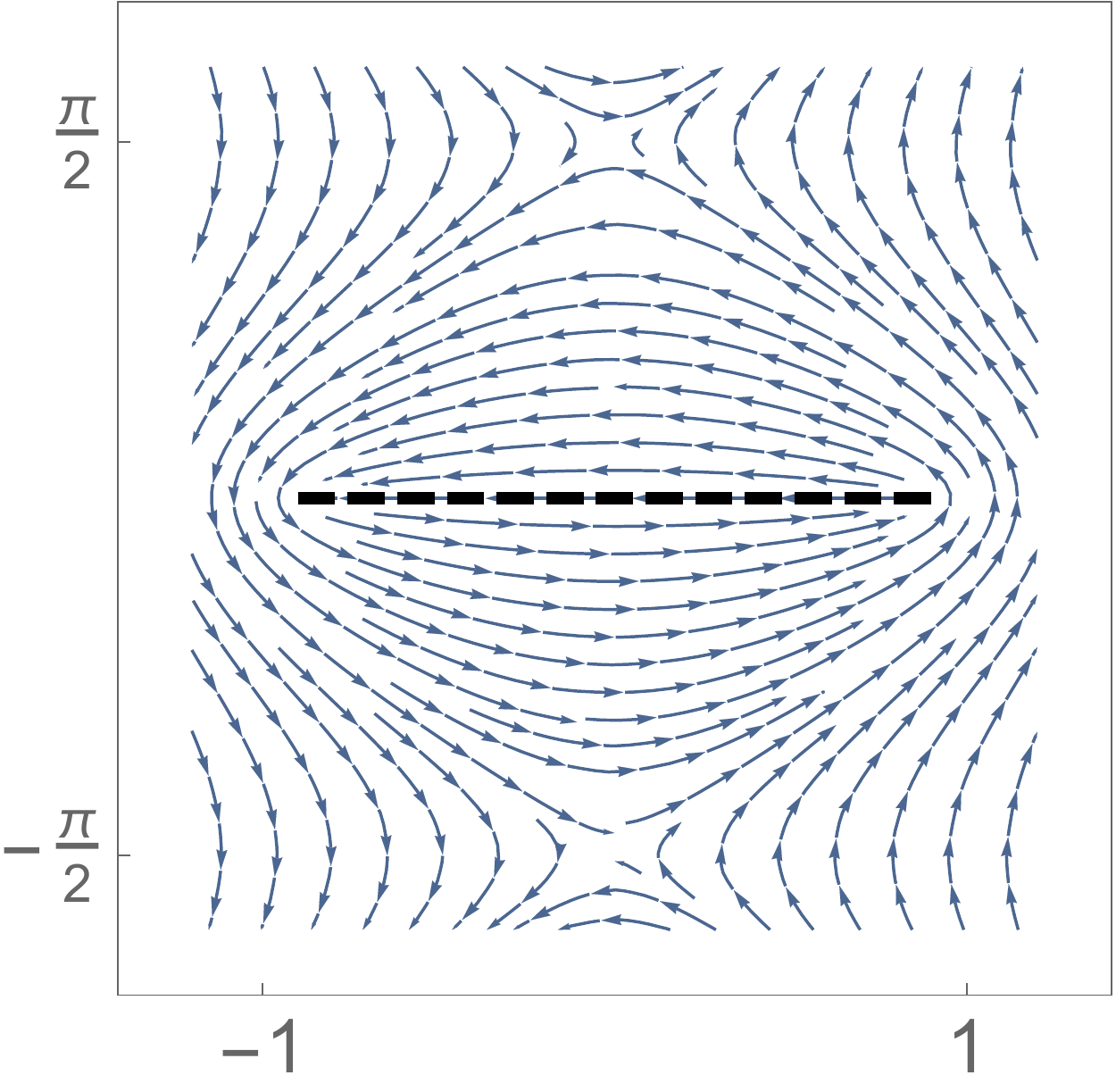}
\caption{\label{tanh-streamplots} Streamplots in the complex $q$--plane for the Sauter pulse instantons. Top to bottom: $c\gamma^2=3,0.1,0,-0.1,-0.5$. Arrows show the direction of the velocity $\dot{q}$.}
\end{figure}

We now turn to the instanton behaviour as a function of $c$.  A convenient way to visualise the transition between timelike and spacelike inhomogeneities is to plot the streamlines of the instanton velocity $\dot{q}$, using (\ref{SECH-VEL}), as a vector field in the complex $q$-plane. (This method is well suited for studying instantons in more general field shapes as we do not need to solve the equations of motion to obtain the stream plots.) Fig.~\ref{tanh-streamplots} shows the streamlines for $c\gamma^2=3,0.1,0,-0.1,-0.5$.  The branch cut which connects the turning points $\dot{q}=0$ and which is encircled by the worldline instantons is highlighted. For each value of $c\gamma^2$, the largest possible instanton is that which brushes the poles in the potential; the streamlines to the left and right do not form periodic loops\footnote{As the chosen field is periodic in the imaginary direction, so too are the streamlines; parts of the instantons in adjacent periods can be seen at the upper and lower edges of the plots, separated by the poles.}. 

For timelike inhomogeneities, $c>0$, the branch cut lies along the imaginary axis. From (\ref{psech})--(\ref{nsech}), periodic instanton solutions exist for $\bar\gamma$ arbitrarily large or small.  For larger $c\bar\gamma^2$ the turning points approach the poles, but recede from them as $c\bar\gamma^2$ decreases, and the length of the branch cut decreases.

For $c<0$ the field inhomogeneity is spacelike, and the branch cut circulated by $q(\tau)$ lies along the real axis. The cut grows without limit as $-c\bar\gamma^2$ increases, and at $-c\bar\gamma^2=1$ extends all the way to infinity and cannot be circulated. There can be no periodic solutions for $-c\bar\gamma^2\geq 1 $ as can be confirmed from~(\ref{psech}): we must have $\sqrt{|c|} \bar\gamma<1$ as otherwise the argument of the square brackets in~(\ref{psech}) hits ({\it independent} of the value of $\tau_0$) the branch cut of $\sinh^{-1}$ and the instanton fails to be periodic. 

\begin{figure}[t!]
	\centering	\includegraphics[width=0.7\columnwidth]{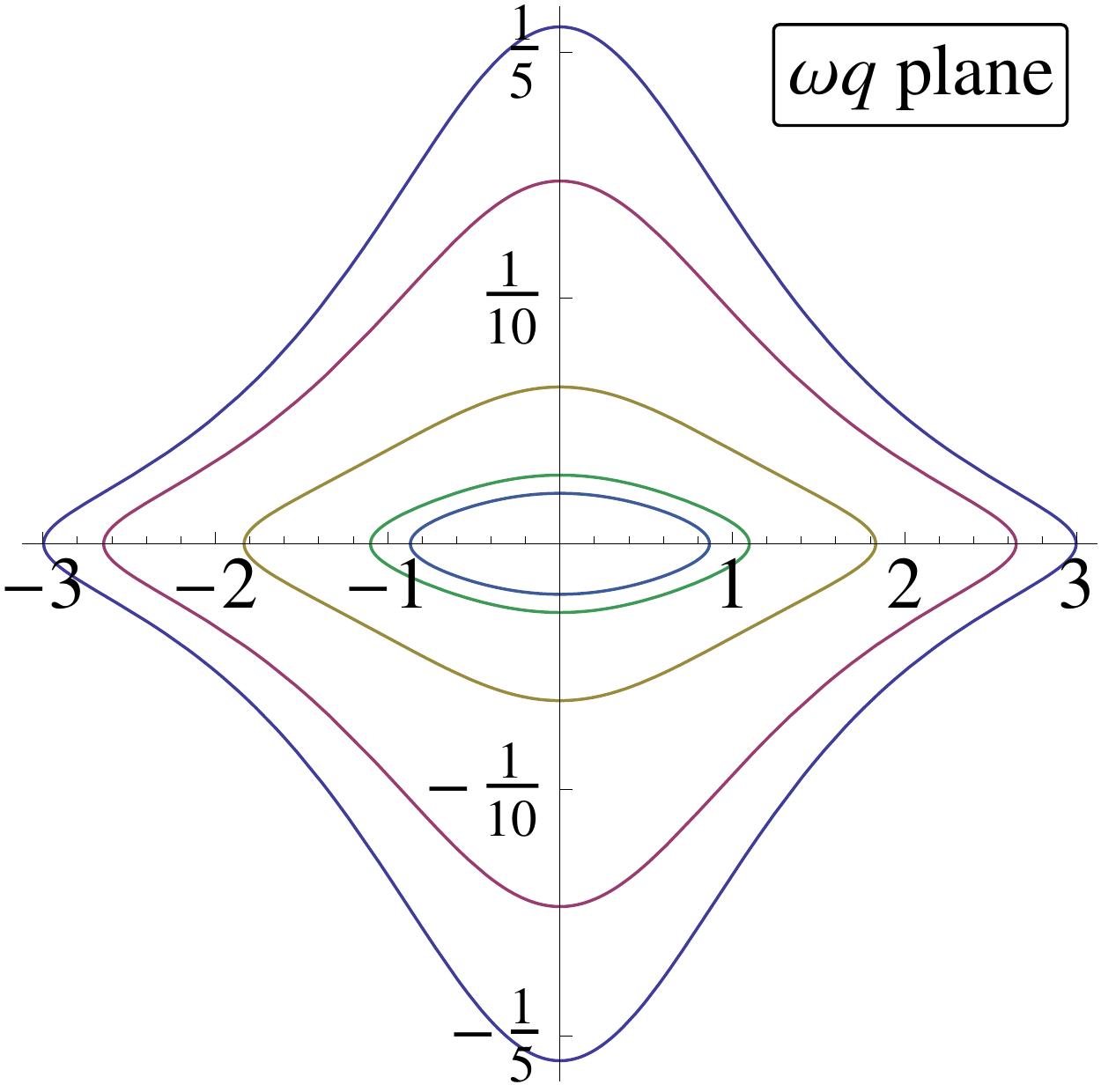}
	\caption{\label{FIG:CC} Complex instantons for $c=-1/2$, i.e.~an electric field depending on a spacelike coordinate $E(t\sqrt{3}+z)$. A fixed $\tau_0 = i/300$ is chosen, and $\gamma$ varies up to the cutoff value of $\gamma=\sqrt{2}$ from (\ref{villkor}), beyond which there are no periodic solutions.}
\end{figure}

This parameter constraint is, again for $a=1$ and reinserting the electron mass,
\be\label{villkor}
	\sqrt{|c|}\, \frac{m\omega}{eE} < 1 \;.
\ee
Hence an electric field with spatial inhomogeneity and given strength must have a certain frequency, or width, in order to be able to produce pairs. This `criticality' is already well-known and has been studied explicitly for the case $c=-1$~\cite{Nikishov:1970br,Gies:2005bz,Dunne:2006st}. Here we see the simple generalisation to the case that the field depends on a spacelike combination of $t$ and $z$. Note that the manifestation of this physics in the worldline formalism is not that the instantons fail to be real for $\sqrt{|c|}\gamma>1$, but that they fail to be {\it periodic}. This is demonstrated explicitly in Fig.~\ref{FIG:CC}, in which we plot the instantons for the field $E(t\sqrt{3}+z)$.


We will find below that the constraint (\ref{villkor}) must also be satisfied in order to have nonzero $\text{Im }\Gamma$ -- for discussions of the connection between the existence of instantons and the possibility of pair production, see~\cite{Dunne:2006st,Dietrich:2007vw}.

As $c$ increases from negative values toward $0$, i.e.~as the field inhomogeneities become `less' spacelike, the constraint on $\bar\gamma$ weakens. The constraint vanishes precisely when $c=0$, the point at which the inhomogeneities become lightlike, and beyond which they are timelike again.

Fig.~\ref{tanh-streamplots} shows that as $c\to 0$, from either above or below, the turning points coalesce and the branch cut disappears. What remains is a (simple) pole; the instantons in the lightlike case are therefore contractable to points, whereas the instantons in all other cases $c\not=0$ are contractable only around {\it extended} objects, namely the branch cuts. This is what singles out the case of lightlike inhomogeneities. It is in just this case, where we lose the extended structure within the instanton loops, that the effective action $\Gamma$ agrees with the locally constant field approximation~\cite{Tomaras:2000ag,Tomaras:2001vs,Fried:2001ur}. That the instantons are contractable to points is consistent with the direct calculation of (\ref{start Gamma}) in~\cite{Ilderton:2014mla}, in which the Minkowski space loops contributing to $\Gamma$ were seen to localise in surfaces of constant lightfront time.

To recover a given instanton solution of the lightlike case $c=0$ from the instantons with $c\not=0$ we must scale some of the instanton parameters with $c$. Consider (\ref{psech}) in the limit $c\to 0$. If we want to obtain a finite sized $q$ in this limit then we must scale $\tau_0$ such that either $\{e^{i2\pi\tau_0}\to\infty,e^{-i2\pi\tau_0}\to0\}$ or vice versa. The limit of (\ref{psech}) is then
\be
	q=\frac{1}{\omega}\sinh^{-1}\big( r\, e^{2\pi i\tau}\big) \;,
\ee  
in which $r$ is a constant; these are indeed the solutions found in~\cite{Ilderton:2015lsa}. We will show below that we do not need to perform any extra rescaling in the final result for the pair production probability, i.e.~the observable, which will have a smooth limit as $c\to 0$.

\section{Calculation of the effective action}\label{SECT:GAMMA}

We proceed to calculate $\Gamma$ in the combined fields (\ref{gauge val}) and (\ref{gauge val2}) describing three-component electric fields and two-component magnetic fields depending on~$q$. We allow for more general field shapes than the symmetric fields ($E(q) = E(-q)$) previously considered. 

\subsection{Classical action}
Using the equations of motion, the classical action of an instanton solution is
\be\label{SS}
	S= \frac{T}{2}(1-a^2) + \frac{1}{c} \int\limits_0^1\!\ud\tau\ \frac{1}{T}{\dot q}^2 - \frac{s}{c}  \int\limits_0^1\!\ud\tau\ \dot{q}(A_\para-p_\para)\;.
\ee
Now we observe that all integrals over $\tau\in [0,1]$ can be expressed as complex contour integrals over the instantons themselves by trading $\ud\tau$ for $\ud q$. 
Consider the final term in the above which vanishes for the commonly studied cases of $t$-dependent or $z$-dependent fields, i.e.~$c=\pm1$ and $s=0$. If $q$ were real we would write this term as a total derivative and drop it due to periodicity. If we instead write it as a contour integral over $q(\tau)$ it becomes
\be
	- \frac{s}{c}  \oint\!\ud q\ (A_\para(q) - p_\para)  \;,
\ee
which will indeed vanish provided the instanton does not circulate poles of the potential. We will assume this in what follows. Turning to the remainder of the action, we introduce the function $G$ to compactify notation:
\be\begin{split}
	G(a^2,p) &= \int\limits_0^{1}\ud\tau\, {\dot{q}}^2 /(c T^2) \\
	&= \frac{1}{c}\oint\ud q\sqrt{ca^2+(A-p)_\para^2+c(A-p)_\LCperp^2} \;,
\end{split}
\ee
where the same branch is chosen as for the velocity, from~(\ref{dq}). The remainder of the action can then be written
\be\label{detta}
	S=\frac{T}{2}(1-a^2)+G(a^2,p) \;.
\ee
As the classical action is an integral over the complex instanton itself, it must be invariant under contour deformation of the instanton, as the integrand in $G$ is analytic away from the branch cut and poles. Changes in the parameter $\tau_0$ introduced in Sect.~\ref{SECT:INST} simply correspond to those deformations for which the instanton remains a solution to the classical equations of motion~\cite{Ilderton:2015lsa}. Hence instantons with different $\tau_0$ give the same contribution to the classical action; this is a complex generalisation of the reparameterisation invariance which allows us to choose an arbitrary real $\tau_0$. This is useful, see below, in defining integrals which would otherwise apparently be divergent. This extends the results of~\cite{Ilderton:2015lsa} to both time-dependent and position dependent electric fields.  

We will find that all terms in the effective action $\Gamma$ can be expressed in terms of $G$ and and its derivatives
\be
	G_0:=\partial_{a^2}G \qquad G_i:=\partial_{p_i}G \qquad \text{etc.,}
\ee
where $i=\{\parallel,\perp\}$. $G$ is closely related to $g$ in~\cite{Dunne:2006st}, except that we include more arguments corresponding to the momenta $p$. Essentially the same function appears in the WKB treatment of pair production in~\cite{Strobel:2013vza}.


Changing variables from $\tau$ to $q$, we can also write down two periodicity constraints
\bea
	1 &=& \langle 1 \rangle=\oint\frac{\ud q}{\dot{q}} \;, \\
\label{PAA}
	\abar_i &=& \langle A_i \rangle  \implies 0=\oint\ud q\frac{A_i-\abar_i}{\dot{q}} \;,
\eea
which (implicitly) determine $a$ and $\abar$, as previously found for real instantons in various fields~\cite{Marinov,Bulanov:2003aj,CPL,Schneider:2014mla}. In terms of $G$ these constraints become
\bea
\label{condition-1}
	1 &=& \frac{2}{T}G_0(a^2(T),p(T)) \;, \\
\label{condition-p}
	0 &=&  G_i(a^2(T),p(T)) \;.
\eea
\subsection{The fluctuation determinant}
We now turn to the contribution of fluctuations about the instantons found above. We will be brief and highlight only differences between this and existing, similar, calculations in the literature.

The second variation of the action in (\ref{start Gamma}) with respect to a fluctuation $\delta x$ around an instanton solution is
\be\label{second variation 3c}
S_2=\int\limits_0^1\frac{1}{2T}\begin{pmatrix} \delta q & \delta d & \delta x^1 & \delta x^2 \end{pmatrix}\Lambda\begin{pmatrix}\delta q \\ \delta d \\ \delta x^1 \\ \delta x^2 \end{pmatrix} \;,
\ee
where
\be\begin{split}
\Lambda &=\begin{pmatrix} -c\partial^2 & s\partial^2 &0&0 \\ s\partial^2 & c\partial^2 &0&0 \\ 0&0&\partial^2 &0 \\ 0&0&0&\partial^2 \end{pmatrix} \\
&+\begin{pmatrix} \frac{\dot{d}}{\dot{q}}\partial\Big(\frac{c\ddot{d}+s\ddot{q}}{\dot{q}}\Big)+\frac{\dot{x}^\LCperp}{\dot{q}}\partial\Big(\frac{\ddot{x}^\LCperp}{\dot{q}}\Big) & \frac{c\ddot{d}+s\ddot{q}}{\dot{q}}\partial & \frac{\ddot{x}^1}{\dot{q}}\partial & \frac{\ddot{x}^2}{\dot{q}}\partial \\ -\partial\frac{c\ddot{d}+s\ddot{q}}{\dot{q}} &0&0&0 \\ -\partial\frac{\ddot{x}^1}{\dot{q}} &0&0&0 \\ -\partial\frac{\ddot{x}^2}{\dot{q}} &0&0&0 \end{pmatrix} \;.
\end{split}
\ee
The $\delta x$ integral is Gaussian and gives the determinant of $\Lambda$, which can be computed with the Gelfand-Yaglom method as in \cite{Dunne:2006st}. Zero modes are avoided by means of Dirichlet boundary conditions, $\delta x^\mu(0)=\delta x^\mu(1)=0$, and an integral over the initial instanton position $x^\mu(0)$. 

The determinant is given by\footnote{This method applies to the ratio of two determinants. We have divided by the free determinant, which is equal to one in our conventions.} $\det\Lambda=\det{}^\alpha\phi^\beta(1)$, where ${}^\alpha\phi$ are the four solutions of the Jacobi equation
\be\label{Jacobii 3c}
\Lambda\phi=0
\ee 
satisfying
\be\label{initialphii 3c}
{}^\alpha\phi(0)=0 \qquad {}^\alpha\dot{\phi}^\beta(0)=\delta^{\alpha\beta} \qquad \alpha,\beta=1,2,3,4 \;,
\ee
where the upper left indices distinguish the four solutions while the upper right denote vector components. We find the solutions by multiplying (\ref{Jacobii 3c}) with $\begin{pmatrix} \dot{q}&\dot{d}&\dot{x}^1&\dot{x}^2 \end{pmatrix}$ and $\begin{pmatrix} 0 & \delta_{ij} \end{pmatrix}$, $j=1,2,3$, and expanding in terms of the trivial solutions,
\be
{}^\alpha\phi={}^\alpha h\begin{pmatrix} \dot{q} \\ \dot{d} \\ \dot{x}^1 \\ \dot{x}^2 \end{pmatrix}+\begin{pmatrix} 0 \\ {}^\alpha d^d \\ {}^\alpha d^1 \\ {}^\alpha d^2 \end{pmatrix} \;.
\ee
Solutions can be expressed in terms of the constants ${}^\alpha k^{1,2,3,4}$
\be
{}^\alpha\dot{h}=\frac{1}{\dot{q}^2}\big(-{}^\alpha k^4c+{}^\alpha k^3(s\dot{q}+c\dot{n})+{}^\alpha k^\LCperp c\dot{x}^\LCperp\big) \;,
\ee
and
\be
{}^\alpha\dot{d}^d=\frac{1}{c}\big({}^\alpha k^3-{}^\alpha\dot{h}(s\dot{q}+c\dot{d})\big) \qquad {}^\alpha\dot{d}^\LCperp={}^\alpha k^\LCperp-{}^\alpha\dot{h}\dot{x}^\LCperp  \;.
\ee
It is now straightforward to impose the boundary conditions \eqref{initialphii 3c}, which leads to long and not particularly illuminating expressions for ${}^\alpha\phi(\tau)$. For $\tau=1$ though all these expressions can be written in terms of derivatives of $G$. We encounter for example terms such as
\be
	\int\limits_0^1\frac{1}{\dot{q}^2}=-\frac{4}{cT^3}G_{00} \;,
\ee
where the left hand side appears, at first sight, divergent because of the turning points $\dot{q}=0$. However, these singularities are avoided by the complex instantons which circulate the turning points, so that $\dot{q}\ne0$ along the whole contour. We can then, if we choose, deform the contour in $G$ down to the branch cut without encountering any divergences. Turning point singularities are also avoided with such contours in the WKB/phase-integral formalism~\cite{Kim:2007pm,Bender:1977dr}. 
After some algebra we find that the determinant can be compactly written as \footnote{In our approximation we consider only the instantons with lowest turning number, $n=1$. As a result the Morse index is fixed by the requirement that the probability be positive.}
\be\label{det Lambda}
\begin{split}
\det\Lambda &=\det{}^\alpha\phi^\beta(1) \\
&=-\frac{4\dot{q}_0^2}{T^6}\Big(G_{00}-G_{0i}G_{ij}^{-1}G_{0j}\Big)\det G_{ij} \;,
\end{split}
\ee
where $i=\{\parallel,\perp\}$. This reduces to the determinant in \cite{Strobel:2013vza} for fields with $G_{0i}=0$ (e.g. for $p=0$) and $c=1$, and to the determinant in \cite{Dunne:2006st} for one-component fields with $c=\pm1$ and $p=0$. In general though $G_{0i}$ is nonzero. This is in a sense a moot point, though, as everything in the round brackets in (\ref{det Lambda}) will soon cancel against a similar term coming from the proper time integral.   

We turn to the $x^\mu(0)$ integrals; those over $d(0)$ and $x_\LCperp(0)$ give volume factors. Including the square root of the determinant in~(\ref{det Lambda}) we have the integral
\be\label{start-int}
	\int\ud q(0)\frac{1}{\dot{q}(0)} \;.
\ee
If the $q_0$-contour were the same as in (\ref{condition-1}), then (\ref{start-int}) would equal one. Instead (\ref{start-int}) contributes a factor of~$1/2$. This is shown for real Euclidean instantons in~\cite{Dunne:2006st}. To see it in the complex case, we propose a contour that starts (ends) on the real axis to the left (right) of the branch cut, and passes either above or below it. In the lightfront limit, the branch cut shrinks to a pole and the usual $i\epsilon$-prescription, $m^2\to m^2-i\epsilon$ in (\ref{S1}), would instruct us whether to go above or below. As the instantons are symmetric under reflection about the real axis, the integral (\ref{start-int}) contributes a factor of $1/2$. 

Finally we perform the proper time integral in a saddle point approximation.  When varying the action (\ref{detta}) in order to obtain the saddle point equations, one must remember that~$a$ and $p$ can depend on $T$.  Here the two conditions \eqref{condition-1} and \eqref{condition-p} for $a$ and $p$ become useful and greatly simplify the proper time derivative:
\be\begin{split}
	\frac{\ud S}{\ud T} &=\frac{\partial S}{\partial T}+\frac{\ud a^2}{\ud T}\Big(-\frac{T}{2}+G_0\Big)+\frac{\ud p_i}{\ud T}G_i \\
	&=\frac{1}{2}(1-a^2) \;.
\end{split}
\ee
Hence the saddle point is determined simply by setting $a^2=1$, as suggested earlier. At the saddle point we have
\be
	S\to G(1,p) \;, \quad T \to 2G_0(1,p) \;, \quad G_i(1,p)=0 \;,
\ee
where the final equation determines $p$ in terms of $c$ and other field parameters. We also need the second variation of the action with respect to $T$. After some simplification, again using \eqref{condition-1} and \eqref{condition-p}, this becomes
\be\label{d2S}
\frac{\ud^2S}{\ud T}=-\frac{1}{4}\Big(G_{00}-G_{0i}G_{ij}^{-1}G_{0j}\Big)^{-1} \;,
\ee
in which we recognise the same factor as appears in~\eqref{det Lambda}, with $a^2=1$.
\subsection{Final result}

Collecting terms from the proper time and path integrals we find
\be\label{final3c}
\text{Im }\Gamma=V\frac{\sqrt{2\pi}}{8\pi^2}\frac{e^{-iG}}{\sqrt{\det(iG_{ij})}} \;,
\ee
where
\be
	G(1,p)=\frac{1}{c}\oint\ud q\sqrt{c+(A-p)_\para^2+c(A-p)_\LCperp^2}
\ee
and $p_{\para,\LCperp}$ are determined by
\be\label{average3}
	G_i(1,p)=0 \;.
\ee
The large round brackets in \eqref{det Lambda} and (\ref{d2S}) have canceled, a  simplification that might have been expected given Gutzwiller's trace formula~\cite{Dietrich:2007vw}. That $iG$ is real follows from the fact that the contour in $G$ can be chosen along an instanton, and that the integrand is the velocity up to a factor of $i$.  As noted in~\cite{Dunne:2006st} we do not need to find the instantons to evaluate (\ref{final3c}); $q$ is now only an integration variable and we can choose any contour that circulates the branch between the turning points where the square root in~$G$ vanishes.

The final result \eqref{final3c} may also be obtained by using the saddle point method for a WKB momentum integral
\be
	\text{Im }\Gamma=\frac{V}{2}\int\frac{\ud^3p}{(2\pi)^3}e^{-iG} \;.
\ee
It was shown in~\cite{Kim:2003qp} that for longitudinal fields with antisymmetric monotonic potentials (with $c=-1$) the wordline instanton results in~\cite{Dunne:2006st} agree with WKB/phase-integral results with saddle point at zero canonical momentum ($p=0$). Equivalence between the worldline formalism and WKB was also shown in \cite{Strobel:2013vza} for three-components fields with $p=0$ and $c=1$. Now we have shown that the two formalisms are equivalent also for more general field shapes with $p\ne0$ and for $-1< c <1$.

Consider now the transition {\it between} the cases of time-dependent and position-dependent fields, as $c$ changes sign. In the limit $c\to 0$ the transverse fields become a plane wave and drop out. Further, all square roots drop out and the branch points and cut coalesce into a pole that can be used to perform the contour integrals in $G$ and its derivatives using Cauchy's residue theorem, as in~\cite{Ilderton:2015lsa}. The instantons are deformable to the poles, i.e.~to points in this limit. It is the contraction of the branch cut which leads to the localisation of the instantons in the lightfront limit. The residue for (\ref{condition-p}) becomes proportional to the derivative of the electric field, implying~\cite{Ilderton:2015lsa}
\be\label{pbar}
	\qquad \abar_\para=A_\para(\bar{q}) \quad \text{and} \quad {E^3}'(\bar{q})=0 \;,
\ee 
so that the instanton circles an extrema of the electric field.

In the prefactor calculation, the factor in the large round brackets in the fluctuation determinant (\ref{det Lambda}) vanishes, which signals the presence of zero modes and suggests that, although the final results are simple on the lightfront, the calculation can be more subtle if it is performed {\it at} $c=0$ from the outset. In this sense nonzero $c$ acts as a regulator (which was one of the original motivations for introducing coordinates interpolating between instant-form time $t$ and front-form $t+z$ in~\cite{Hornbostel:1991qj,Ji:2001xd,Ji:2012ux}) and allows us to complete the calculation in~\cite{Ilderton:2015lsa} by extending the localisation seen in the instantons to the whole effective action. In the limit $c\to 0$ the effective action becomes
\be\label{LC}
	\text{Im }\Gamma\to\frac{V_3}{16\pi^3}E^2\sqrt{\frac{2E^2}{-E''}}e^{-\pi/E} \;,
\ee
where, from (\ref{pbar}), the field is evaluated at its maximum. \eqref{LC} agrees with the locally constant field approximation.

\section{Timelike vs. spacelike inhomogeneities}\label{SECT:PRAT}
We now study examples of the effective action (\ref{final3c}), beginning with purely longitudinal fields, $A_\LCperp=0$.

\subsection{Longitudinal fields}

For this case we have $G_{ij}\propto\delta_{ij}$ and (\ref{final3c}) reduces to
\be\label{final long}
	\text{Im } \Gamma=\frac{\sqrt{2\pi}}{32\pi^2}V_3\frac{e^{-iG}}{iG_0\sqrt{-iG_{00}/c}} \;.
\ee
To generalise to spinor QED, we simply have to multiply (\ref{final long}) with a factor of $2$, as can be shown in the same way as in~\cite{Dunne:2005sx,Dumlu:2011cc}. 

To make the dependence on $c$ more obvious we will take the usual case of antisymmetric, monotonically increasing potentials $A(q)$ with $\abar=0$; in short the same field shapes as in~\cite{Dunne:2006st} but now depending on our general coordinate $q$. Let $A(q)=f(\omega q)/\gamma$, in which $f$ is a dimensionless, monotonically increasing shape function with $-1<f<1$. Changing variable from $q$ to $y$ defined by $f(\omega q)=\sqrt{-c\gamma^2}y$ we find
\be\label{Gg}
	G(1,0)=-i\frac{\pi}{eE_0}g(c\gamma^2) \;,
\ee
where $g$ is~\cite{Dunne:2006st}
\be\label{ggg}
	g(c\gamma^2):=\frac{2}{\pi}\int\limits_{-1}^1\ud y\frac{\sqrt{1-y^2}}{f'} \;,
\ee
with $f'$ expressed in terms of $f^2=-c\gamma^2y^2$. We note briefly for this class of fields that, changing variables from $q$ to $A$ as in~\cite{Dunne:2005sx,Dumlu:2011cc}, one can show that
\be\label{Tpole}
	T=-\frac{2\pi i}{\langle eE\rangle} \;,
\ee
which is the same relation as was found in~\cite{Ilderton:2015lsa} for lightfront-time dependent fields ($c=0$). In the case of constant fields, for which $\langle eE\rangle \to eE_0$, (\ref{Tpole}) gives the location of the poles in the proper-time integral, the residues from which generate the imaginary part of the effective action, i.e.~give pair production~\cite{CS}.

Returning to the effective action, we substitute (\ref{ggg}) into (\ref{final long}) to obtain
\be\label{final2}
	\text{Im }\Gamma = \frac{V_3\sqrt{2}}{32\pi^3}\frac{(eE_0)^{3/2}}{\gamma}
	\frac{e^{-\frac{\pi}{eE_0}g(\zeta)} }{\partial_\zeta[\zeta g(\zeta)]\sqrt{-\partial_\zeta^2[\zeta g(\zeta)]}}
	\bigg|_{\zeta=c\gamma^2} \;.
\ee
This generalises the result in~\cite{Dunne:2006st}: the difference is that a factor of $c$ now multiplies the adiabaticity parameter squared,~$\gamma^2$. In \cite{Dunne:2006st} it was noted that the result for $A(x)$ is obtained from that of $A(t)$ by $\gamma^2\to-\gamma^2$, which in our interpolating coordinates corresponds to taking $c$ from $1$ all the way to $-1$. Importantly, the effective action is a smooth function for all $c$, and in particular is continuous as $c$ goes through $0$.

\begin{figure}[t!]
	\centering\includegraphics[width=\columnwidth]{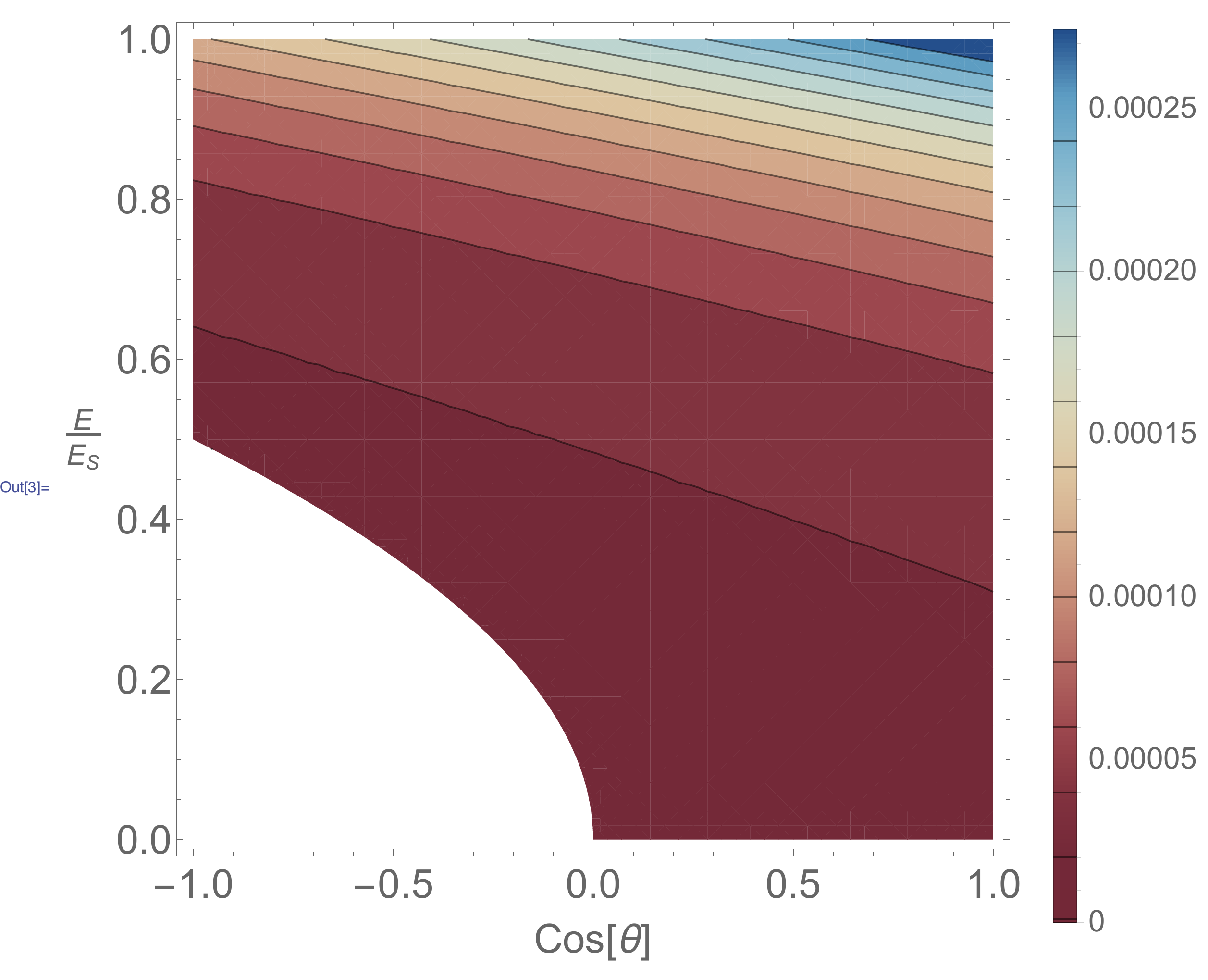}
	\caption{\label{c-E-kontur} Contour plot of the effective action $\text{Im }\Gamma/V^3$ as in (\ref{att plotta}), in units of the electron mass. In order to show structure, the frequency is fixed to the high value of $\omega=m/2$. For $c<0$, when the field inhomogeneity is spacelike, the electric field strength must obey $E_0 /E_S > \sqrt{|c|}/2$ in order to be capable of producing pairs. We see that to achieve a given pair production probability it is preferable for the field to have temporal rather than spatial inhomogeneities.}
\end{figure}

\begin{figure}[t!]
	\centering\includegraphics[width=\columnwidth]{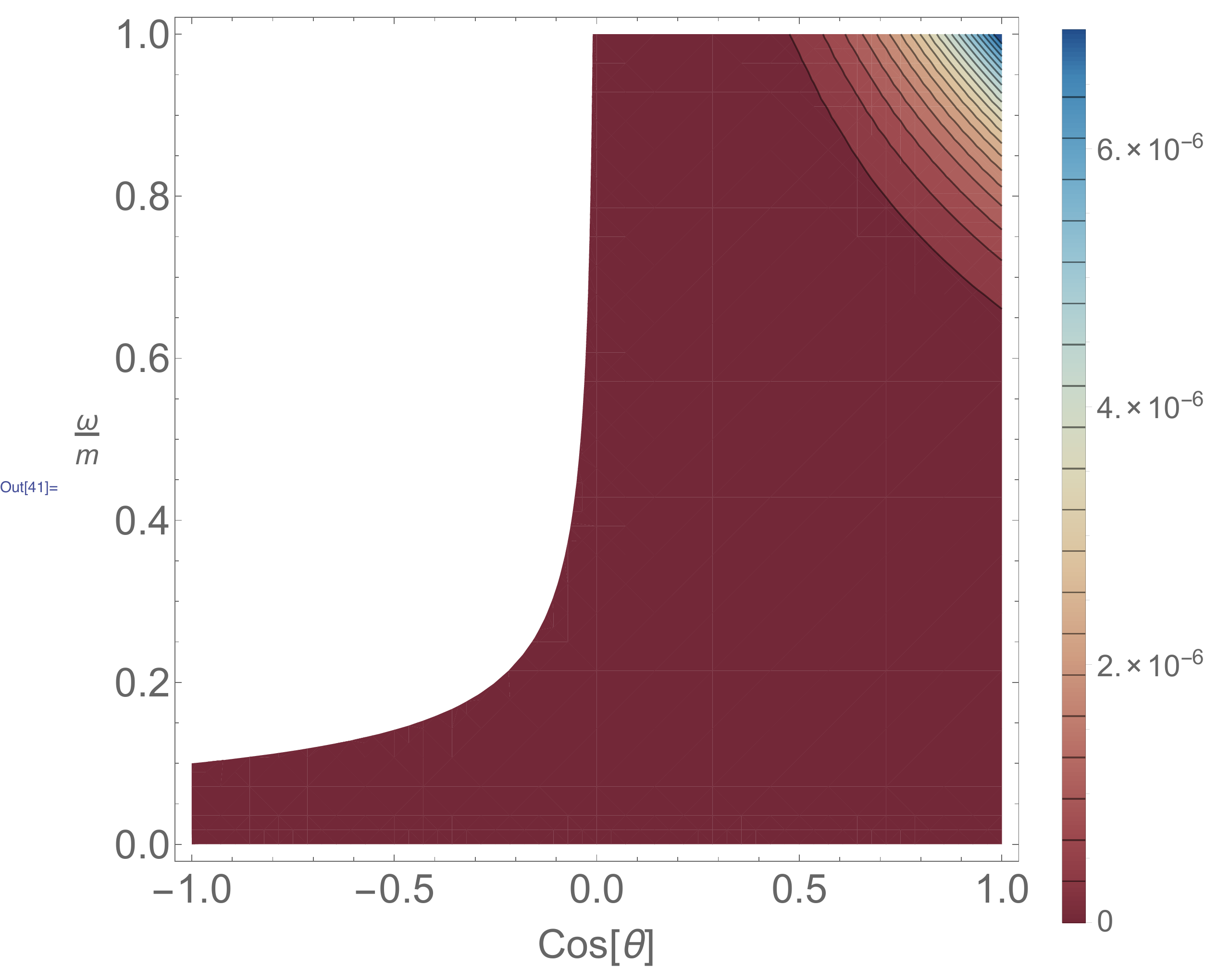}
	\caption{\label{c-E-kontur-2} Contour plot of the effective action $\text{Im }\Gamma/V^3$ as in (\ref{att plotta}), in units of the electron mass. Here $E=E_S/10$, and the critical curve is at $100c (\omega/m)^2=-1$. }
\end{figure}

To illustrate, take the Sauter pulse (\ref{Sauter}). Reinstating the electron mass and writing $\mathcal{E}=eE_0/m^2 = E_0/E_S$, the ratio of the peak field to the Schwinger field, one finds
\be\label{att plotta}
	\frac{\text{Im }\Gamma}{m^3V_3}=\frac{\mathcal{E}^{3/2}}{16\pi^3\gamma}(1+c\gamma^2)^{5/4}\exp \bigg( -\frac{\pi}{\mathcal{E}}\frac{2}{1+\sqrt{1+c\gamma^2}} \bigg)\;,
\ee
see also~\cite{Dunne:2006st} and, for higher-order corrections,~\cite{Kim:2007pm}. Fig.~\ref{c-E-kontur} shows a contour plot of the effective action (\ref{att plotta}) as a function of $c$ and fieldstrength $\mathcal{E}$ at fixed frequency. We see that the probability increases from zero beyond the critical curve
\be\label{cg2}
	c\gamma^2=-1 \;.
\ee
This curve connects the known critical point $\gamma=1$ for the case of spatially inhomogeneous fields $E(z)$ to $\gamma=\infty$ at $c=0$, where the field dependency becomes lightlike, $E(z)\to E(t+z)$. For $c>0$ on the other hand, in which the field dependency is timelike, there are no critical points. The same behaviour is seen in Fig.~\ref{c-E-kontur-2} where we plot the effective action as a function of $c$ and frequency $\omega/m$ at fixed field strength. Hence the lightfront limit $c=0$ separates systems with critical points from those without. It is already known that for $q=t$ the pair production probability is larger than given by the locally constant approximation, while for $q=z$ it is smaller, and for the lightlike case, $q=x^\LCp$, equal; the contours in Fig.~\ref{c-E-kontur} confirm that (at fixed field strength and frequency) any timelike dependence yields a higher pair production rate than any spacelike dependence.

It is worth saying that for the near future's physically realisable parameters, i.e.~electric fields delivered by intense laser systems of optical frequency $\omega\sim1$eV and field strengths of order at most $E_S/100$, (\ref{att plotta}) varies only imperceptibly, at the order of 1 part in $10^9$, as we interpolate between spacelike and timelike inhomogeneities. Hence the biggest distinction between the spacelike and timelike cases comes from the different volume factors appearing which, after regularisation, go roughly like the Rayleigh range for $c=+1$ and the pulse length for $c=-1$. (The remaining area factor, common to both volumes, is the focal spot size.)

There is a simple argument which allows us to check~(\ref{final2}). Let us make the frequency/wavelength explicit in the argument of the field, writing $E\equiv E(\omega q)$. For, e.g.~$0\le\theta<\pi/2$ we boost with velocity $v=\tan(\theta/2)$, parallel to the field, such that the interpolating coordinate $q$ becomes proportional to time in the new frame, $q=\sqrt{c}t'$. In this new frame we have a time dependent electric field $E(\sqrt{c}\omega t')$~\cite{Becker:1976pp,Nikishov:2002ja,Bulanov:2003aj}, with frequency scaling with $c$ as
\be\label{kc}
	\omega'=\sqrt{c} \,\omega \;.
\ee 
Further, the volume $V^d$ in the $d$-direction is (for $|t'|\ll V$) related to the volume in the $x'$-direction by $V'=\sqrt{c}V^d$. Hence, by {\it partially} undoing our rotation with a Lorentz transformation, we can check (\ref{final long}) by rescaling parameters in the result of~\cite{Dunne:2006st}.

For $c<0$ we can similarly boost to a frame with a~$z'$ dependent field and frequency (or rather wave vector) scaling as $\omega'=\sqrt{-c}\omega$. If there are transverse components the same Lorentz transformation simplifies them: for $c>0$ ($c<0$) the magnetic (electric) components vanish after the boost. However, the magnitude of the transverse components also scales with $\sqrt{\pm c}$, which means that the corresponding adiabaticity parameters are independent of $c$.

This argument also informs the lightlike limit $c\to 0$. In the final expression for $\Gamma$ we can see that $c\to0$ takes the adiabacity to zero and, comparing with (\ref{kc}), is effectively a zero {\it frequency} limit. This is again consistent with the lightlike case agreeing with the locally constant field approximation.

\subsection{Longitudinal vs. transverse components}
\begin{figure}[t!]
	\centering\includegraphics[width=\columnwidth]{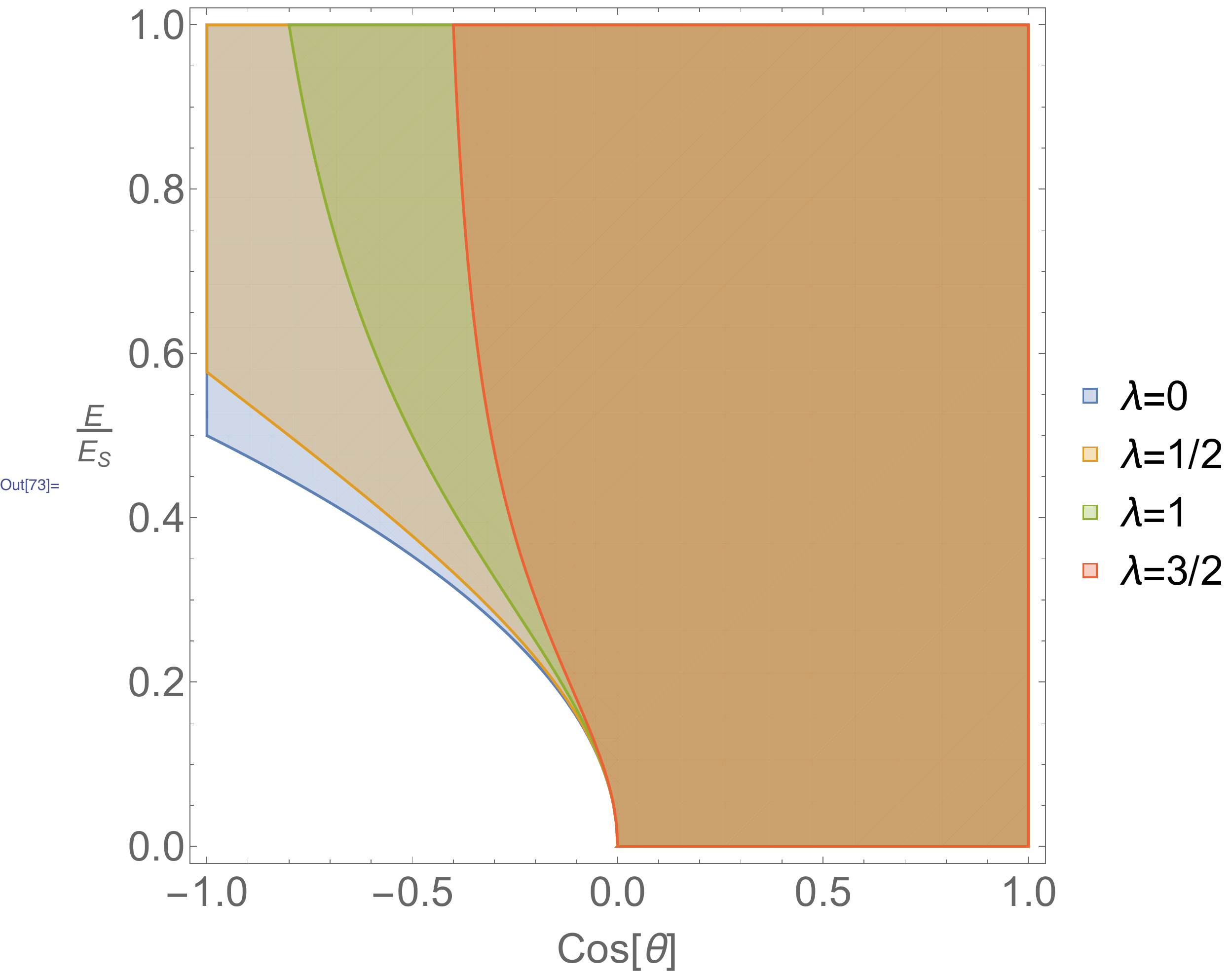}
	\caption{\label{FIG:LAMBDA} Effect of a transverse field on the criticality condition, compare (\ref{nykritt}) with (\ref{cg2}). Parameters as in Fig.~\ref{c-E-kontur}. For fields with spacelike inhomogeneities $c<0$, turning on transverse electric and magnetic fields {\it reduces} the parameter region in which pair production is possible.}
\end{figure}
Consider now adding an additional electric field component which points in the transverse direction (along with the magnetic component) as in~(\ref{EB-trans}). We take $A_1\not=0$. Let this and the longitudinal field component have the same shape but different magnitudes,
\be
	A_1(q) = \lambda\, A_\para(q) \;,
\ee
with $\lambda$ dimensionless. Assume again that $A_\para = f/\gamma$ is antisymmetric and monotonic, then we have
\be
	G= \frac{1}{c} \oint\!\ud q\, \sqrt{c+(1+c\lambda^2)A_\para^2} \;, 
\ee
which, comparing with the purely longitudinal case, simply amounts to rescaling $\gamma^2$ by factor $1/(1+c\lambda^2)$. The criticality condition then becomes
\be\label{nykritt}
	\frac{c\gamma^2}{1+c \lambda^2} > -1 \;,
\ee
provided $1+c\lambda^2>0$ is also obeyed (as otherwise the invariant ${\bf E}^2-{\bf B}^2$ becomes negative and, ${\bf E}.{\bf B}$=0, there is no pair production.) The constraint (\ref{nykritt}) reduces to (\ref{cg2}) for purely longitudinal fields, $\lambda=0$; the two constraints are compared in Fig.~\ref{FIG:LAMBDA}. For $c<0$ we see that the effect of the transverse fields is to reduce the volume of parameter space in which pair production is possible.

\subsection{Two-component `rotating' fields}
We turn now to a two-component transverse field
\be\label{rot-def}
	A_\LCperp(q)=\frac{1}{\gamma}\{\cos\omega q,\sin\omega q\} \;,
\ee
which for $c=1$ is a time-dependent rotating field, and for $c=0$ a transverse (monochromatic) plane wave. (For $c< 0$ one can boost to a frame where the field becomes purely magnetic and there is no pair production.) For the time-dependent case the effective action was studied in~\cite{Marinov,Bulanov:2003aj,Strobel:2013vza} using WKB, in~\cite{Blinne:2013via} using the Wigner formalism, and the Euclidean instantons found in~\cite{CPL}.  We will therefore be brief here, only using this as an example to show that the complex structures above exist in other fields. We believe though that this is the first calculation of $\Gamma$ including the prefactor contribution, using the worldline approach.

We begin with the instantons. Fulfilling the periodicity constraint (\ref{PAA}) or (\ref{average3}) is essential here. Given the form of the field (\ref{rot-def}), we make the following ansatz for $p_\LCperp$:
\be\label{rot-ave}
	\abar_\LCperp=\frac{\rho}{\gamma} \{\cos\omega \bar{q},\sin\omega \bar{q}\} \;,
\ee
where $\rho :=\gamma |\abar_\LCperp|$ is the magnitude of the transverse momentum, and $\bar{q}$ is some average which, as we will shortly confirm, may be freely chosen. The constraint which determines $\rho$ as a function of $\gamma^2$ is given in~\cite{CPL}. This constraint may be rewritten in terms of the complete elliptic integrals ${\bf K}$ and ${\bf E}$ (see~\cite[\S17.3]{Abram}) as
\bea
	\label{conditionh}
	0 &=& \frac{4}{\pi\sqrt{\rho}}\Big[(1+\rho){\bf K}(-\sigma^2)-2{\bf E}(-\sigma^2)\Big] \;,
\eea
in which $\sigma^2 := (\gamma^2+(\rho-1)^2)/(4\rho)$. Numerical solution of (\ref{conditionh}), or estimation using the approximations in~\cite{CPL}, shows that $\rho>1$, which is larger than would be possible if the instanton were real. To see the reason for this let $\bar{q}=0$ and deform the instanton so that $q(\tau)$ follows a straight line between the turning points, recall~(\ref{dq}). The line is along the imaginary axis, so while one component of the average is zero $\langle\sin \omega q\rangle=0$ the other is $\langle\cos \omega q \rangle=\langle\cosh i\omega q\rangle >1$.

\begin{figure}[t!]
\includegraphics[width=0.3\columnwidth]{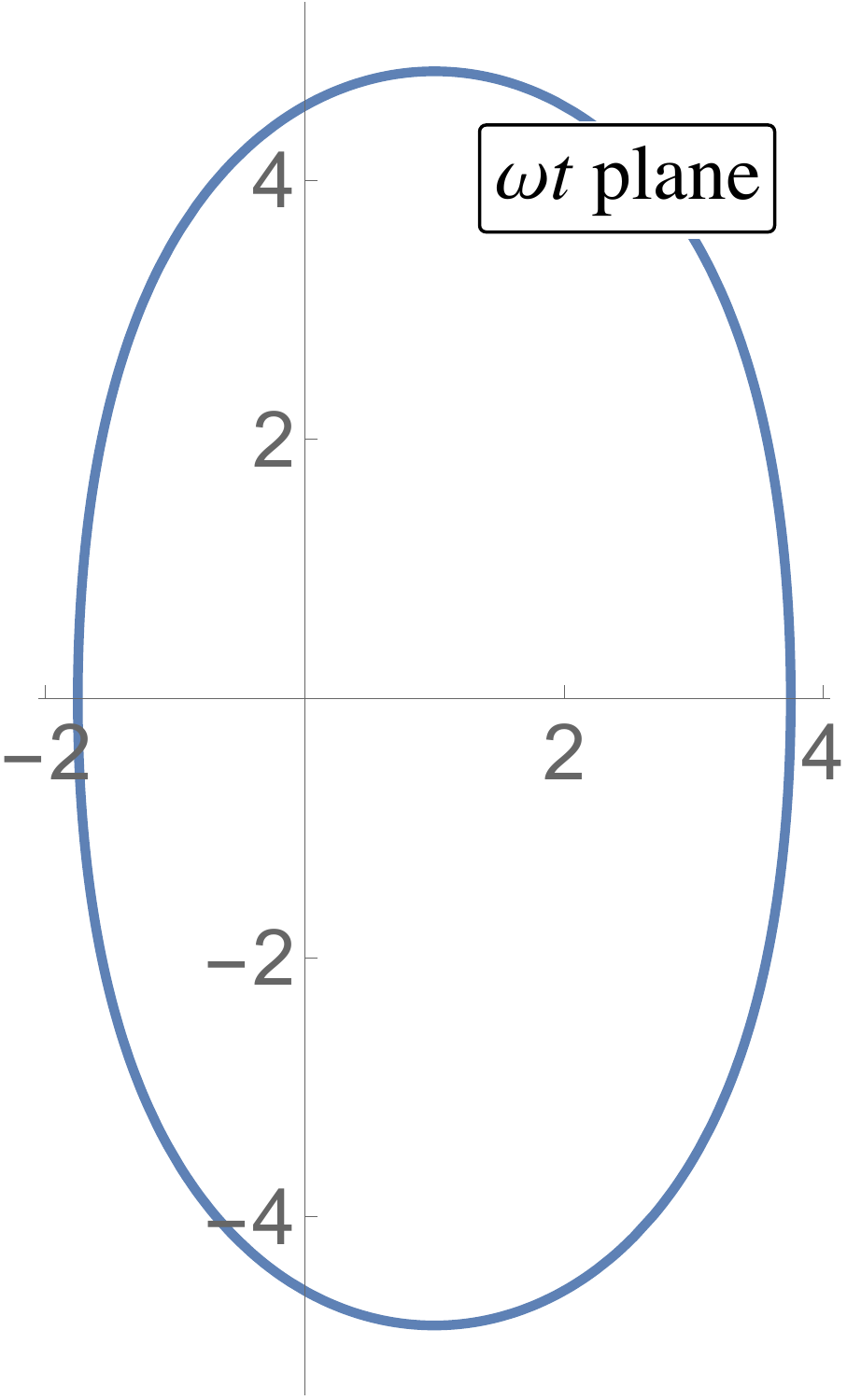}
\includegraphics[width=0.3\columnwidth]{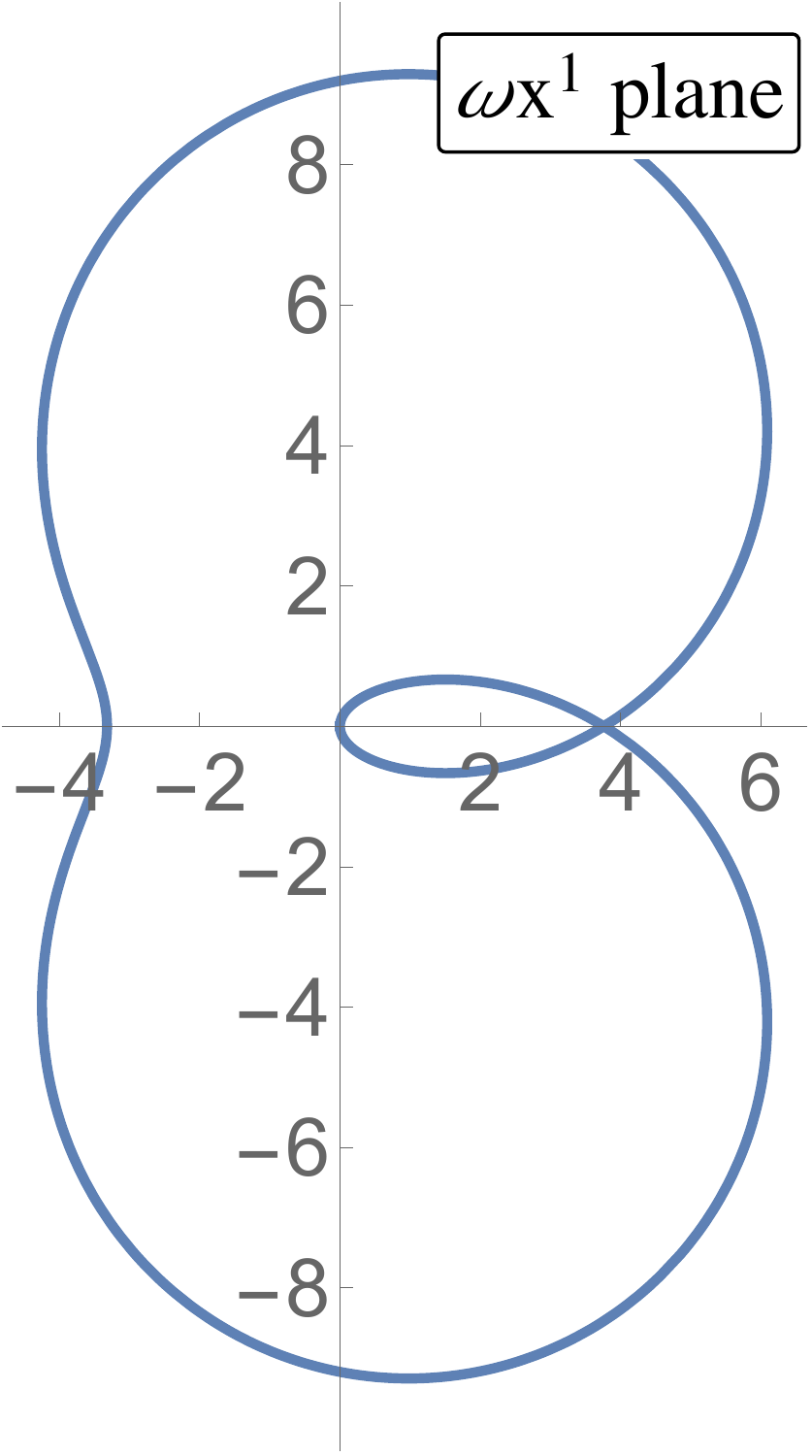}
\includegraphics[width=0.35\columnwidth]{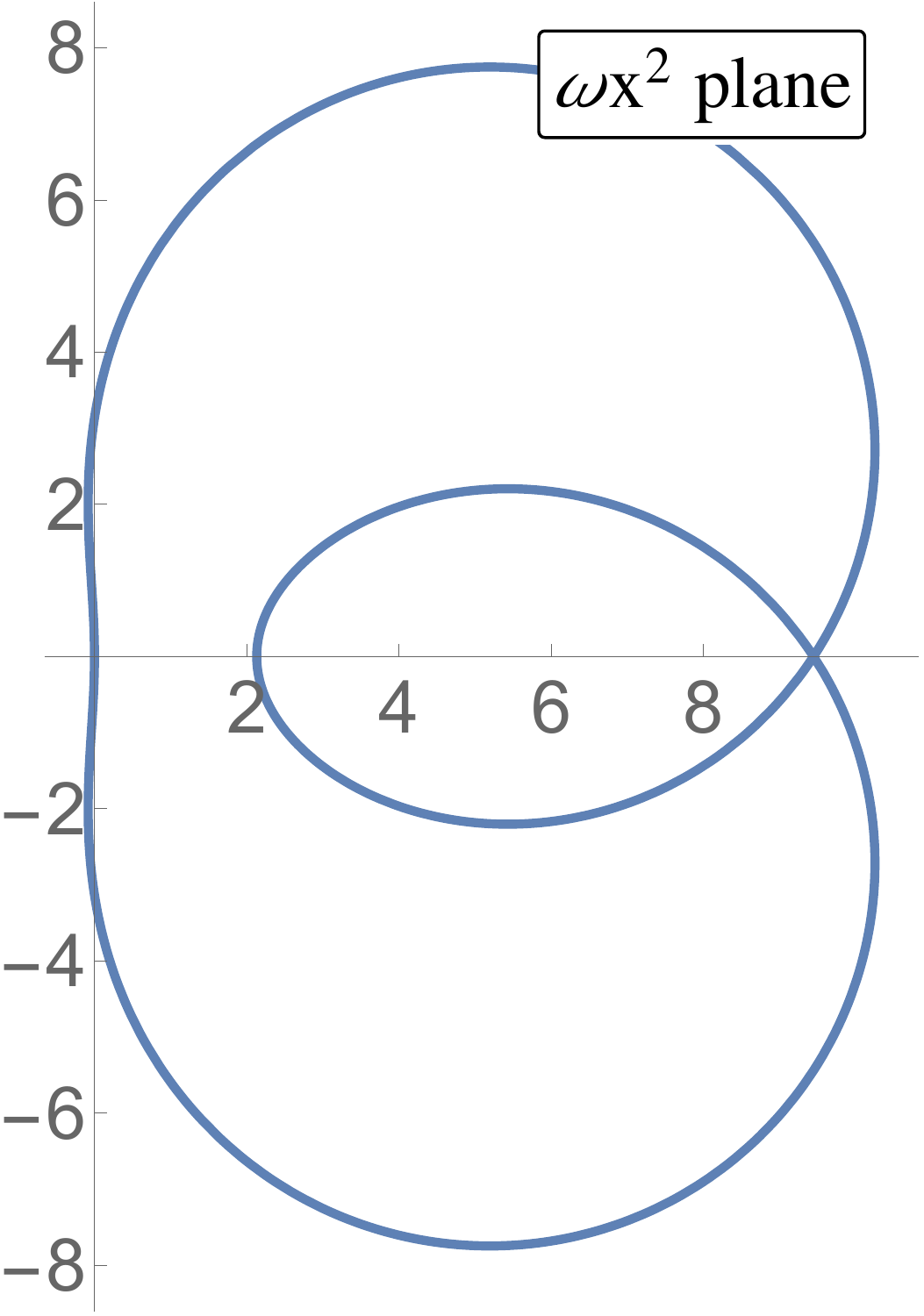}
\caption{\label{self-intersecting} Complex instantons $\omega t$, $\omega x_1$ and $\omega x_2$ for $c=\gamma=1$, with $\omega\bar{t}=1$ and initial condition $\omega t(0)=-1.75$.
}
\end{figure}

In~\cite{CPL} further conditions were imposed on the instantons to ensure reality, but these are not needed. This is confirmed in Fig.~\ref{self-intersecting}, where the instantons for $c=1$ are found by numerical integration of (\ref{Lorentz}). A real $\bar{t}$ and a real staring point $t(0)$ are chosen, and indeed the solutions are still complex. We have confirmed that periodic solutions are only found if (\ref{conditionh}) is fulfilled. While $t(\tau)$ remains a simple closed curve, the instantons $x^\LCperp(\tau)$ can self-intersect.

\begin{figure}[t!]
	\centering\includegraphics[width=\columnwidth]{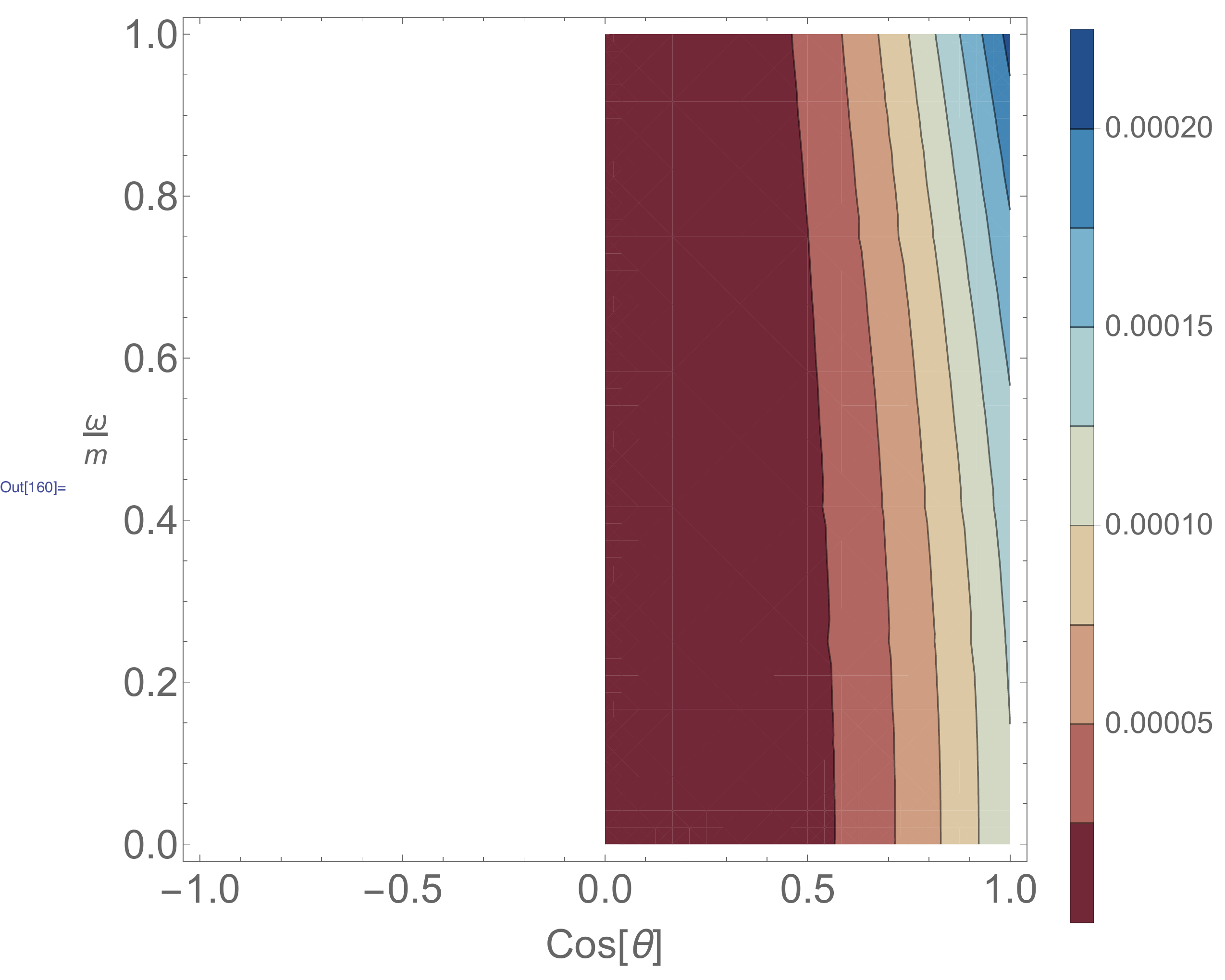}
	\caption{\label{FIG:GAMMA-ROT} Contour plot of the effective action (\ref{Gammah}) with $E_0=E_S$. There is no pair production for $c<0$.}
\end{figure}

The classical action of an instanton, $G$, can be written
\be\begin{split}
	iG &=\frac{\pi}{\sqrt{c} E_0\gamma^2}h(\gamma^2,\rho) \;, 
\end{split}
\ee
where $h$ again depends on elliptic functions\footnote{It is interesting to compare (\ref{hKE}) with Eq.~$3.51$ in~\cite{Dunne:2006st}, which gives $g(\gamma^2)$ for a longitudinal oscillating field, $E_0\cos\omega t$; the only difference, after using various elliptic function identities, is the factors of $\rho$, which are absent in the longitudinal case, and an overall factor of $2$.}:
\be\label{hKE}
		h(\gamma^2,\rho) = \frac{8}{\pi}\sqrt{\gamma^2+(\rho+1)^2}\Big[{\bf K}-{\bf E}\Big]\Big(\frac{\gamma^2+(\rho-1)^2}{\gamma^2+(\rho+1)^2}\Big) \;,
\ee
as also appears in the WKB calculation of~\cite{Strobel:2013vza}. Due to the symmetry of the field $G$ does not depend on $\bar{q}$. In the calculation of the effective action this leads to a zero mode, and the determinant of $G_{ij}$ appearing in (\ref{final3c}) vanishes. To separate out the zero mode into a volume factor we follow~\cite[\S 39.4]{ZinnJustin:2002ru}. First write the determinant as a momentum integral,
\be\label{average-integral}
	\frac{e^{-iG}}{\sqrt{-\det iG_{ij}}}=\frac{1}{\sqrt{-iG_{33}}}\int\frac{\ud\abar_\LCperp}{2\pi}e^{-iG} \;,
\ee
as holds to lowest order in the semiclassical regime, and change variables from $p_\LCperp$ to $\rho$ and $\omega\bar{q}$; the $\rho$ integral is performed with the saddle point method and the $\bar{q}$ integral gives a volume factor $\omega V_q$, or $2\pi$ per period of the rotating field. The result is
\be
	\frac{e^{-iG}}{\sqrt{-\det iG_{ij}}}=\frac{i}{\sqrt{2\pi}}\frac{V_tE}{\gamma}\frac{\rho}{\sqrt{iG_{33}iG_{\rho\rho}}}e^{-iG} \;,
\ee
and the effective action becomes
\be\label{Gammah}
	\text{Im }\Gamma = \frac{V_4 c (eE_0)^2\rho}{8\pi^3 \sqrt{2h_1h_{22}}} \exp\bigg(\displaystyle -\frac{\pi}{e\sqrt{c}E_0}\frac{h}{\gamma^2}\bigg) \;.
\ee
extending the results of~\cite{Bulanov:2003aj,CPL} to $c<1$. The dependence of $\Gamma$ on $c$ is plotted in Fig.~\ref{FIG:GAMMA-ROT}.

\section{Conclusions}\label{SECT:CONC}
We have used a coordinate rotation to investigate Schwinger production in backgrounds which interpolate between time-dependent, homogeneous electric fields and inhomogeneous, static electric fields. This allowed us to examine the transition between Lorentz-inequivalent spacetime dependencies. For all field dependencies we found that the instanton contribution to the effective action was given by a complex contour integral over the instanton itself, with the physics of pair production being encoded in the branch cuts circulated by the instantons.

Note that the existence of critical points, beyond which there is no pair production, is not related to the {\it reality} of the instantons. Instead we have seen that critical points arise simply when the instantons fail to be {\it periodic}.

Being complex contours, the instantons can be freely deformed (Cauchy's integral theorem) around the branch cuts, without changing their contribution to the effective action. A striking property of the instantons is that they make this symmetry manifest: the freedom to choose $\tau_0$ in the instanton solutions represents those deformations for which the instanton remains a solution to the equations of motion. This was previously found for the case of lightlike field dependencies~\cite{Ilderton:2015lsa}, but now we have seen that it holds more generally, for both spacelike and timelike dependencies, and for a range of field configurations including longitudinal fields and two-component rotating fields.

For all timelike and spacelike inhomogeneities, the instantons are deformable only down to a branch cut, and are therefore fundamentally extended objects. We found though that the limit of {\it lightlike} coordinate dependence corresponded to a vanishing field frequency scale, in which the branch circulated by the instantons contracted to a pole. The instantons in this case are contractable to points, i.e.~are equivalent to pointlike objects, and it is in just this limit that the effective action is given simply by the locally constant approximation. We have recovered this result by rotating our field dependence from timelike to spacelike, and the effective action is a continuous function of the interpolating parameter.

The freedom to deform the instantons, even away from solutions to the equations of motion, seems consistent with the observation in~\cite{Dunne:2006st} that the explicit form of the instantons is not strictly needed in order to calculate the semiclassical approximation to the pair production probability. For the example of the Sauter pulse we have seen that, at~fixed field strength and frequency, the probability is higher for any timelike dependence than for any spacelike dependence.
 
We have considered fields with one dominant maximum and consequently negligible interference effects. Such effects have been studied in~\cite{Dumlu:2011cc} using a phase-space worldline approach, and it would be interesting to study the cases covered there using our formalism.

\acknowledgments
A.I.~and G.T.~are supported by The Swedish Research Council, contract 2011-4221.

\end{document}